\documentclass[useAMS]{mn2e}
\usepackage{times}
\usepackage{rotate}
\usepackage{psfig}
\usepackage{amssymb}
\title[The Deep Chandra Survey of the Groth Strip - II.]{The Deep
  Chandra Survey of the Groth Strip - II. optical identification of
  the X-ray sources} 
\author[Georgakakis et al. ] {A. Georgakakis$^{1}$\thanks{Marie Curie fellow}, K. Nandra$^{1}$, E. S. Laird$^{1}$, S. Gwyn$^2$,
C. C. Steidel$^{3}$, V. L. Sarajedini$^{4}$, \\\\ 
{\rm \LARGE P. Barmby$^{5}$, S. M. Faber$^{6}$, A. L. Coil$^{7}$,
M. C. Cooper$^8$, M. Davis$^{8,9}$, J. A. Newman$^{10}$
} 
  \\ \\
  $^1$Astrophysics Group, Blackett Laboratory, Imperial College, Prince
  Consort Rd , London SW7 2BZ, UK\\
  $^2$Department of Physics and Astronomy, University of Victoria,
  P.O. Box 3055, Victoria, BC V8W 3P6, Canada.\\  
  $^3$California Institute of Technology, Pasadena, CA 91125, USA\\
  $^4$University of Florida, Department of Astronomy, Gainesville, FL
  32611, USA\\
  $^5$Harvard-Smithsonian Center for Astrophysics, 60 Garden Street,
  Mail Stop 65, Cambridge, MA 02138, USA\\
  $^6$UCO/Lick Observatory and Department of Astronomy and
  Astrophysics, University of California, Santa Cruz, CA
  95064, USA\\
  $^7$Steward Observatory, University of Arizona, 933 N. Cherry Ave.,
  Tucson, AZ 85721-0065, USA\\
  $^8$Department of Astronomy, University of California, Berkeley, CA 
94720\\
  $^9$Department of Physics, University of California, Berkeley, CA 
94720\\
  $^{10}$Hubble Fellow; Institute for Nuclear and Particle Astrophysics, 
Lawrence Berkeley National Laboratory, Berkeley, CA 94720 
}
\begin{document}
\maketitle  

\begin{abstract}
In this paper we discuss the optical and X-ray spectral properties of
the sources detected in a single 200\,ks {\it Chandra} pointing in the
Groth-Westphal Strip region. A wealth of optical photometric and
spectroscopic data are available in this field providing optical
identifications and redshift determinations for the X-ray
population. The optical photometry and spectroscopy used here are
primarily from the DEEP2 survey with additional redshifts 
obtained from the literature. These are complemented with the deeper
($r\approx26$\,mag) multi-waveband data ($ugriz$) from the Canada
France  Hawaii Legacy Survey to estimate photometric redshifts and to
optically identify sources fainter than the DEEP2 magnitude limit 
($R_{AB}\approx24.5$\,mag). We focus our study on the 2-10\,keV
selected sample comprising 97 sources to the limit $\approx  8 \times 
10^{-16}\rm \, erg \, s^{-1} \, cm^{-2}$, this being the most complete
in terms of optical identification rate (86\%) and redshift 
determination fraction (63\%; both spectroscopic and photometric). We
first construct the redshift distribution of the sample which shows a
peak at $z\approx1$. This is in broad agreement with models where less
luminous AGNs evolve out to $z\approx1$ with powerful QSOs peaking at
higher redshift, $z\approx2$. Evolution similar to that of broad-line
QSOs applied to the entire AGN population (both type-I and II) does
not fit the data. We also explore the {\it observed} $N_H$
distribution of the sample and estimate a fraction of obscured
AGN ($N_H \rm > 10^{22} \, cm^{-2}$) of $48\pm9$ per cent. This is
found to be consistent with both a  luminosity dependent {\it
  intrinsic} $N_H$ distribution, where less luminous systems comprise
a higher fraction of type-II AGNs and models with a fixed ratio 2:1
between type-I and II AGNs. We further compare our results with those
obtained in deeper and shallower surveys. We argue that a luminosity
dependent parametrisation of the intrinsic $N_H$ distribution is
required to account for the fraction of obscured  
AGN observed in different samples over a wide range of fluxes.   
\end{abstract}

\section{Introduction}\label{sec_intro}

In the last few years the study of the diffuse X-ray background (XRB)
has witnessed significant observational progress allowing detailed
comparison with model predictions. The ultra-deep {\it Chandra}
surveys in particular, have demonstrated that most of the XRB, at both
soft and hard energies, is resolved into discrete point sources
(Brandt et al. 2001; Giaconni et al. 2002; Alexander et al. 2003), the
vast majority of which are without doubt AGNs. To the first approximation,
this finding has been a huge success for models that reproduce the
spectral properties of the XRB under the zero order assumption that it
originates in a combination of obscured and unobscured AGN
(Comastri et al. 1995; Gilli, Salvati \& Hasinger 2001). Under more
careful examination however, a number of inconsistencies
emerge. Firstly, luminous ($L_X>10^{44} \rm \, erg \, s^{-1}$) heavily 
obscured type-II QSOs at $z\approx1.5-2$, predicted in large numbers
by the models, are scarce in the surveys above.  Secondly, the
redshift peak of the X-ray population lies below $z=1$ in stark
contrast with the model expectation of $z\approx1.5-2$. 

Although these inconsistencies suggest that some revision of the
models is almost certainly required (Hasinger 2003),
observational biases may complicate any interpretation. For example,
about $\approx25$ per cent of the sources in the {\it  Chandra} Deep
Fields (CDF) are optically faint, $R>24$\,mag, rendering optical 
spectroscopy  difficult or even impossible with current technology
(Rosati et al. 2002; Barger et al. 2003). Any information about the
nature of these sources is therefore limited and they are proposed as 
best candidates for heavily obscured AGN (Alexander et al. 2001;
Treister et al. 2004), likely to comprise a fraction the elusive
population of high-$z$ type-II QSOs. Moreover, the small field-of-view
of the CDFs (0.07\,deg$^2$ each) makes them sensitive to cosmic
variance further complicating interpretation of the derived redshift
distribution.  

Wide-area shallower ($\approx 10^{-14} \rm \, erg \, s^{-1} \,
cm^{-2}$) surveys are less affected by the observational biases above
(e.g. Baldi et al. 2002; Kim et al. 2004; Georgantopoulos et
al. 2004). These samples although of key importance, comprise a large 
fraction of unobscured AGNs that are not representative of the sources
responsible for the spectral shape of the XRB ($\Gamma=1.4$;
e.g. Gruber et al. 1999).    

The evidence above suggests that deep surveys with relatively
wide field-of-view are essential to improve our understanding of the
XRB. Observational programs in this direction are already well
underway such as the {\it XMM-Newton} Cosmic Evolution Survey [COSMOS;
$\rm 2\,deg^2$ , $f_X ( \rm 0.5 - 2 \, keV) \approx 5\times 10^{-16} \, erg
  \, s^{-1} \, cm^{-2}$] and the Extended {\it Chandra} Deep Field
South  [E-CDF-S; Lehmer et al. 2005; Virani, Treister \& Urry 2006;
$\rm 0.3 \, deg^2$, $f_X ( \rm 0.5 - 2 \, keV) \approx  1.1 \times
10^{-16} \, erg \, s^{-1} \, cm^{-2}$]. In 
this paper we present  results on a single  200\,ks {\it Chandra}
pointing, which is part of an on-going X-ray survey in the Extended
Groth Strip region, which will eventually cover a total area of about
0.5\,deg$^2$ to the depth above (200\,ks per pointing). This sample,
when completed, will be intermediate in terms of area coverage and
depth to the CDFs and shallower wide-area surveys, minimising any
observational biases affecting the ultra-deep fields and comprising a
large fraction of obscured AGNs responsible of the XRB properties
(Nandra et al. 2005). Moreover, the Extended Groth Strip is targeted
by the largest space and ground-based facilities for  multiwavelength
observations: (i) the DEEP and DEEP2 surveys provide  optical
spectroscopy to the limit $R_{AB} \approx 24$\,mag, (ii) multiwaveband
optical photometry to fainter magnitudes is underway as part of the
Canada France Hawaii Telescope Legacy Survey, (iii) deep imaging and
spectroscopy, independent from the programs above, has been performed
by Steidel et al. (2003) in search for Lyman break galaxies, (iv)
comprehensive imaging with HST/ACS has recently been completed, (v)
{\it Spitzer} mid-IR data are available, (vi) radio observations to
sub-mJy levels have been obtained by Fomalont et al. (1991) with new
much wider VLA observations recently completed, (vii) SCUBA has
observed part of this field to the deep limits of the Canada-UK Deep
Submillimetre survey (Webb et al. 2003). A combination of the X-ray
observations with the mutliwavelength datasets above promises a
breakthrough in the study of the evolution and large scale structure
of AGNs as well as the connection between AGN activity and host galaxy
formation.     

This paper presents the optical and X-ray spectral properties of
the sources detected in the first 200\,ks {\it Chandra} pointing
observed as part of the Extended Groth Strip X-ray survey. This
observation encompasses the original Groth-Westphal Strip region
(Groth et al. 1994). In addition to studying the properties of the
X-ray sources in the context of XRB models, our purpose is to
demonstrate the power of the full $\rm 0.5 \, deg^2$ Extended Groth
Strip {\it Chandra} survey, when completed, for XRB
studies. Throughout this paper we adopt $\rm Ho = 70 \, km \, s^{-1}
\, Mpc^{-1}$, $\rm \Omega_{M} = 0.3$ and $\rm \Omega_{\Lambda} = 0.7$.    

\section{Data}\label{sec_survey}

\subsection{X-ray observations}

The X-ray data used in this paper are from the A03 {\it Chandra}
observations of the original Groth Westphal Strip (GWS), which is part
of the Extended Groth Strip (EGS) region. The total exposure time is
about 190\,ks split into 3 separate integrations obtained at different 
epochs. All 3  observations were obtained with the ACIS-I instrument
($17^{\prime} \times 17^{\prime}$) with a similar roll-angle at the
aimpoint, $\alpha$ = 14:17:43.6, $\delta=$52:28:41.2. A detailed
description of the data reduction, source detection and flux
estimation has been presented by Nandra et al. (2005). 

Briefly, standard reduction methods were applied using the CIAO
version 3.0.1 data analysis software. After merging the individual
observations into a single event file, we constructed images in 4
energy bands 0.5-7.0\,keV (full), 0.5-2.0\,keV (soft), 2.0-7.0\,keV
(hard) and 4.0-7.0\,keV (ultra-hard). Source detection was performed
using a simple but efficient method which is based on pre-selection of
candidate sources using the {\sc wavdetect} task of CIAO followed by
aperture count extraction using the 90 per cent Point Spread Function
(PSF) radius and a local background determination to estimate the
source significance. The final catalogue used in this paper comprises
a total of 158 sources over a total surveyed area of $\rm 0.082 \,
deg^2$ to a Poisson detection  probability threshold 
$<4\times10^{-6}$. Of these sources a total of 155, 121, 97, and 44 
are detected in the full, soft, hard and ultra-hard  bands
respectively. Fluxes are estimated by integrating the net counts
within an aperture corresponding to the 70 per cent encircled energy
radius at the position of the source. The counts in the full, soft,
hard and ultra-hard bands are converted into fluxes in standard bands,
0.5-10, 0.5-2, 2-10 and 5-10\,keV respectively. The limiting flux in
each of these bands is estimated  $\rm 3.5 \times 10^{-15}$, $\rm 1.1
\times 10^{-16}$, $\rm 8.2 \times 10^{-16}$ and $\rm 1.4 \times
10^{-15} \rm \, erg \, s^{-1} \, cm^{-2}$ respectively.   

\subsection{Optical photometry}
The main photometric catalogue used in this paper for the optical
identification of the X-ray sources is the DEEP2 survey of the EGS
that also overlaps with the original GWS field.

The DEEP2 survey photometric data were obtained at the
Canada-France-Hawaii Telescope (CFHT) using the $\rm 12k \times 8k$
pixel CCD  mosaic camera  providing a $\rm 0.70 \times 0.47\, deg^2$
field of view per pointing. The observations were performed  in the
$B$, $R$ and $I$ filters. The data reduction, source detection,
photometric and astrometric calibration as well as the star-galaxy
separation are described in Coil et al. (2004). The pointing that
overlaps with the {\it Chandra} X-ray data used here is nearly complete to
$R_{AB} \approx 24.50$\,mag ($B_{AB} \approx 24.75$,  $I_{AB} \approx
23.5$\,mag). This is shallower than the nominal limit  
of the full DEEP2 EGS survey  ($R_{AB} \approx 24.75$\,mag)
because of poorer  seeing conditions (about 0.95\,arcsec) at the time
of the observations (Coil et al. 2004). The astrometric accuracy of
the photometric catalogue is estimated to be 0.5\,arcsec and is
limited by systematic errors of the USNO-A catalogue used to determine
the  astrometric solution.  

The GWS also overlaps with the ongoing deep synoptic
Canada-France-Hawaii Telescope Legacy Survey (CFHTLS). This project  
uses the wide field imager MegaPrime equipped with the MegaCam CCD
array providing a $\rm 1 \times 1 \, deg^2$  field of view. In this
paper we use the first data obtained as part of the deep synoptic
survey in the $ugriz$ filters. The exposure times in each waveband
range from 1--13 hours (depending on the filter) corresponding to
about 2--9 per cent of the target integration at the completion of the
project. The data reduction, source detection, photometric and
astrometric calibration will be presented in a future paper. In brief
the Elixir package was used for the reduction as well as the initial
photometric and astrometric calibration, which were then refined using
our own routines. The final astrometric uncertainty
is estimated to be about 0.3\,arcsec. The photometric accuracy is found
to be better than 0.05\,mag in all filters while  the completeness
limit in the AB system is $r\approx26$\,mag ($u\approx25.5$,
$g\approx26.0$, $i\approx25.5$, $z\approx25.0$\,mag). Although the
CFHTLS deep synoptic dataset reaches fainter limits that the DEEP2, we
prefer to use the  latter at present as the basic photometric
catalogue because of its homogeneity and the well documented
observational properties of this survey (e.g. Coil et al. 2004; Faber
et al. 2005; Willmer et  al. 2005). We nevertheless use the
multi-waveband photometry of the CFHTLS primarily to estimate
photometric redshifts but also to search for X-ray source optical
counterparts that are fainter than the DEEP2 limit.    

Finally, the GWS has been targeted for deep optical imaging as part of
a larger program searching for Lyman Break galaxies (Steidel et
al. 2003). The observations were performed at the Kitt Peak 4m  Mayall
telescope using the Prime Focus CCD camera ($\rm 14.2 \times 14.2 \,
deg^2$  field of view) in the $Un$, $G$ and $\Re$ filters. Because of
the smaller field-of-view of these observations the outer edges of
X-ray pointing do not overlap with the optical image. A detailed
description is presented by Steidel et al. (2003). The astrometry is
accurate to about 0.4\,arcsec and the photometric internal scatter is
estimated to be better than  0.03\,mag in all filters. These
observations reach a limiting magnitude $\Re_{AB}\approx26$\,mag, 
similar to the CFHTLS. They are used here primarily to estimate
photometric redshifts using the Lyman break selection criteria. As
discussed by Steidel et al. (2003), these methods are very efficient
in identifying galaxies in narrow redshift slices in the range $1.5
\la z \la 3$. 

We note that the $R$-band filters used in the above three datasets are
similar and therefore, there is good  agreement in the estimated
$R$-band magnitudes of the same object among the different surveys. 

\subsection{Optical spectroscopy}

The main source of optical spectroscopy in this study  is the DEEP2
redshift survey. This is an ongoing project that uses the DEIMOS 
spectrograph on the 10\,m Keck II telescope aiming to obtain redshifts
for about 40\,000 galaxies in the range $0.7 \la z \la 1.5$ to a
limiting magnitude $R_{AB}=24.1$\,mag. The spectra are obtained with a
high resolution grating (1200\,l/mm, $R \approx 5000$) and span the
wavelength range $\rm 6500-9100\,\AA$. This spectral window
allows the identification of the  O\,II emission line in the redshift
interval  $0.7 \la z \la 1.4$. Outside this range  the ability to
measure redshifts and hence, the completeness of the DEEP2, drops
significantly. The data reduction was performed using an IDL based
pipeline developed at UC-Berkeley (Cooper et al. 2006) and adapted
from reduction programs created for the SDSS. 

The GWS has also been targeted by a number of spectroscopic programs
(Lilly 1995; Brinchmann et al. 1998; Hopkins et al. 2000; Voght et
al. 2005) that have been compiled into a single database by  Weiner et 
al. (2005\footnote{http://saci.ucolick.org/verdi/public/index.html}). The
entire EGS overlaps with the Sloan Digital Sky Survey (SDSS) and 
therefore spectra for relatively bright galaxies and QSOs are
also available (York et al. 2000).    

In addition to the above surveys Steidel et  al. (2003, 2004)
performed follow-up  multi-slit spectroscopy of the GWS Lyman Break
galaxies as well as some X-ray sources using the LRIS-B on the Keck 
telescopes. The observations used a 300\,line/mm grating blazed at
$\rm 5000\AA$  leading to a dispersion of 2.47\,\AA/pixel, a
wavelength range that included at least the $\rm 4000-7000\,\AA$
regime and a nominal spectral resolution of about $\rm 12.5\AA$. The
total integration time varied between 1.5-3\,h (depending on the
observing conditions), split into 1800\,s sub-exposures
followed by a dither of the telescope in the slit direction. The data
were reduced using a custom package based on IRAF scripts. A total of
10 sources in our sample have redshift measurements from these
observations.  

\section{Optical Identification}

The main catalogue used to optically identify the GWS X-ray sources is
the DEEP2 survey. We first search for systematic offsets between the
astrometric solutions of the X-ray and optical catalogues. A matching
radius of 2\,arcsec is adopted to include only secure optical 
identifications. We also consider X-ray sources with off-axis angles  
$<6$\,arcmin (total of 86) where the {\it Chandra} PSF
is superior with a 90 per cent encircled energy radius of
$\la$4\,arcsec. A total of 50 X-ray sources have optical
identifications brighter than $R_{AB}<24.5$\,mag. We estimate small
systematic offsets of $\delta \rm   RA=-0.23$ and $\delta \rm
DEC=0.37$\,arcsec between the X-ray and optical source
positions. These were then used to align the X-ray source catalogue to
the DEEP2 astrometric solution.    

\begin{figure}
\centerline{\psfig{figure=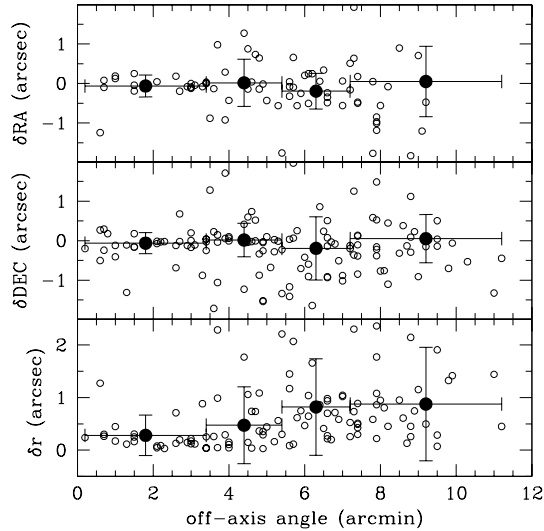,width=3in,angle=0}}
\caption
 {X-ray/optical positional offset in RA (upper panel), DEC (middle
 panel) and total angular distance (lower panel) against X-ray
 off-axis angle. The filled points represent  the 
 mean offset within different off-axis angle bins. The horizontal
 errorbar corresponds to the width of each bin while, the vertical
 errorbar is the $1\sigma$ rms. The width of the bins varies so that
 each of them includes about 25 X-ray/optical pairs. 
 }\label{fig_drra_vs_oaa}
\end{figure}

Next we explore the positional accuracy of the X-ray centroid as a
function of off-axis angle, $\theta$. We match the X-ray and optical
catalogues using an ample 5\,arcsec search radius to account for the
degradation of the PSF at large off-axis angles. Figure
\ref{fig_drra_vs_oaa} plots the positional offset in RA and DEC
between the X-ray and optical source positions against
$\theta$. Reassuringly, the mean X-ray--optical offset is close to
zero at all off-axis angles but the $1\sigma$ rms increases from about
0.5\,arcsec at $\theta\la6$\,arcmin to $\approx1$\,arcsec at larger
off-axis angles. We account for the degradation of the X-ray
positional accuracy by varying the matching radius as a function of
off-axis angle. For  $\theta\le6$\,arcmin we use a radius of
1.5\,arcsec, corresponding to the $3\sigma$ rms scatter
around the mean. For  $\theta>6$\,arcmin the matching radius increases
to 3\,arcsec, the $3\sigma$ rms positional uncertainty at these
off-axis angles.   

The surface density of optical sources to the limit $R_{AB}=24.5$\,mag
is large enough that a substantial fraction of chance associations
is expected within the above radii. We account for this
effect by estimating the Poissonian probability, $P$, that a given
optical counterpart is spurious alignment following the  method of
Downes et al. (1986). Given the surface density of objects brighter
than $m$, $\Sigma(<m)$, the expected number of candidates within $r$
is $\mu=\pi\,r^{2}\,\Sigma(<m)$. Assuming that source positions are
Poissonian, the probability of at least one object brighter than $m$
within radius $r$ is $P=1-\exp(-\mu)$. In practice one has to apply a
cutoff in $P$ to limit the optical identifications to those candidates
that are least likely to be spurious alignments.  

The probability $P$ however, is estimated under the assumption that
the source positions are uniformly distributed within the surveyed
area. For the real clustered distribution of optical sources we assess
the fraction of spurious optical identifications for different
probability cutoffs using mock X-ray catalogues constructed by
randomising the positions of the  X-ray sources within the area
covered by the {\it Chandra} observations. The optical identification
method is performed on the mock catalogues using the same criteria
(e.g. matching radius) as for the real sources. This procedure is
repeated 500 times. When  constructing random X-ray catalogues we
maintain the spatial distribution of the sources due to both
vignetting and real clustering. This is accomplished by applying
offsets in the range 30-60\,arcsec to the X-ray source positions
around their original centroid.  

\begin{figure}
\centerline{\psfig{figure=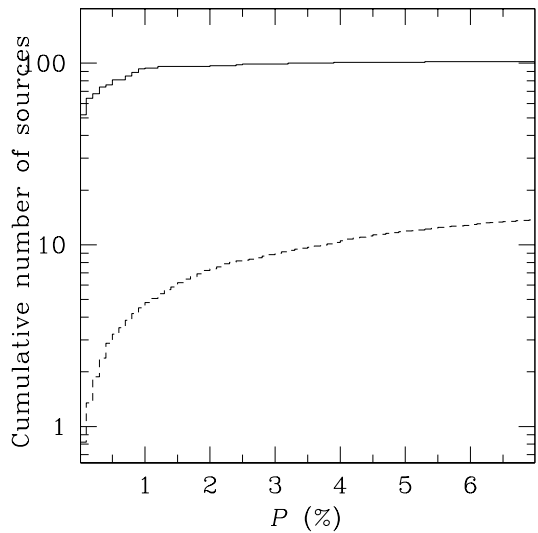,width=3in,angle=0}}
\caption
 {Number of optical identifications as a function of probability
 cutoff $P$ for the full-band sample. The continuous histogram is for
 the real X-ray catalogue. The dashed line corresponds to the mean of
 500 mock X-ray catalogues as described in the text. 
 }\label{fig_prob_vs_numids}
\end{figure}

Figure \ref{fig_prob_vs_numids} plots the cumulative distribution
of optical identifications for the full-band X-ray selected sample
(using the matching radius scheme described above) as a function of
probability cutoff. Also shown in Figure \ref{fig_prob_vs_numids} is
the expected number of spurious counterparts estimated as described
above. This figure shows that the number of optically
identified sources reaches a plateau at $P \approx 1.5$ per cent while
the spurious identification rate further increases with $P$. Based on 
Figure \ref{fig_prob_vs_numids} we adopt a cutoff probability $P<2$
per cent for optical identification in the case of off-axis angles
$<6$\,arcmin. Because of the degradation of the positional accuracy at
larger off-axis angles we relax the probability cutoff to  $P<4$ for
sources with $\theta >6$\,arcmin. This is to minimise the fraction of
missed optical identifications because of the poor X-ray
positions. For an optical source with $R_{AB}=24.5$\,mag the
probabilities $P<2$ and $P<4$ per cent correspond to maximum
separations between the optical and X-ray centroids of about 1 and
1.5\,arcsec respectively. Repeating the simulations above using the
off-axis dependent identification scheme we estimate a spurious
fraction of about 4.5 per cent. The  choice of $P$ is a trade-off
between maximum number of optical counterparts and minimum
contamination rate. Similar results and false identification rates are
obtained for the soft, hard and ultra-hard samples.   

For X-ray sources with no optical identification to the DEEP2
magnitude limit or outside the DEEP2 field-of-view we use the CFHTLS
to search for fainter optical counterparts applying the same selection
criteria described above. Moreover, a number of optically faint X-ray
sources lie in the gaps between the CCDs of the MegaCam mosaic. For
these sources we use the Steidel et al. (2003) deep optical imaging to
search for optical identifications. Considering sources fainter than
the DEEP2 magnitude limit increases the spurious fraction rate by
about 2 per cent. A total of 29 sources are identified with galaxies
from the CFHTLS or the Steidel et al. (2003) survey. Table
\ref{tab_ids} summarises the identification statistics for different
X-ray selected subsamples. Table \ref{tab_optical} presents the
optical properties of the GWS X-ray sources as well as the source of
optical photometry.

\section{Redshift estimation}\label{sec_redshift}
Spectroscopic redshifts are available for a total of 51
sources. These are classified into 3 groups on the basis of their
optical spectroscopic properties (primarily from DEEP1 and DEEP2)
following methods described in Sarajedini et al. (2006): broad
emission-line galaxies, narrow emission-line sources and systems with
absorption lines. For broad emission line AGN we adopt the criterion
$\rm  FWHM>1200\,km\,s^{-1}$. We note that the low S/N ratio of some
of the spectra and the small spectral window of the DEEP2 observations
($\rm 6000-9500\,\AA$) introduce some uncertainty in the
classification scheme above. In addition to the above three groups, a
number of sources in the sample  cannot be classified  because the
optical spectra were not available  for visual inspection.  

For X-ray sources without spectroscopic identification we estimate
photometric redshifts exploiting the multiwaveband photometry 
($ugriz$) of the CFHTLS. Determining photometric redshifts for X-ray
sources is challenging because of the significant, if not dominant,
AGN component contributing to the optical broad-band colours
(e.g. Babbedge et al. 2005; Kitsionas et al. 2005). Recent studies
however, suggest that many moderate luminosity AGN ($\la 10^{44} \rm
erg \, s^{-1}$) as well as obscured X-ray sources have optical
continuum emission that is dominated by stellar light thus, allowing
galaxy templates to be used for photometric redshifts (Barger et
al. 2003; Gandhi et al. 2004; Georgakakis et al. 2004; Georgakakis et
al. 2006).  

We explore this possibility using the photometric redshift code of
Gwyn (2001) based on a standard $\chi^2$ minimisation
method. The galaxy templates are based on those of Coleman, Wu \&
Widman (1980), providing SEDs for 4 main spectral galaxy types (E/S0,
Sbc,  Scd, Im), extended in the UV and IR wavelength regions using the
GISSEL98 code (e.g. Bruzual \& Charlot 1993). These are supplemented
with the starburst SB2 and SB3 spectra from  Kinney et
al. (1996). Having only a small number of SEDs can cause aliasing in
photometric redshifts. Therefore, a new template set has been created
by smoothly interpolating between each of the six original
spectra. This results in a total 51 SEDs from ellipticals (spectral
classification 0) to extreme starbursts (spectral classification 1).   

\begin{figure}
\centerline{\psfig{figure=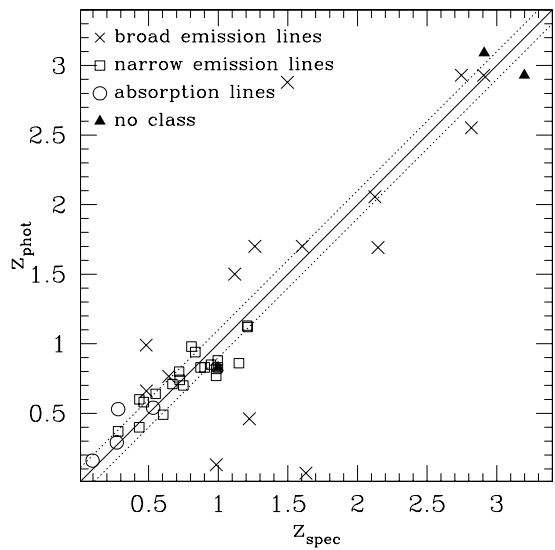,width=3in,angle=0}}
\caption
 { Photometric against spectroscopic redshift estimates for the
 X-ray sources with available spectroscopic observations. 
 Circles are for sources with absorption-line spectra, squares
 correspond to systems with narrow emission-line spectra and crosses
 are broad line AGNs.Triangles are sources with no classification.  
 }\label{fig_zz}
\end{figure}

\begin{figure}
\centerline{\psfig{figure=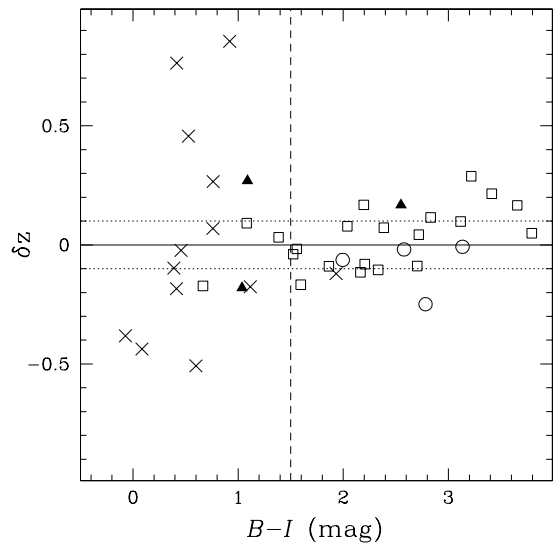,width=3in,angle=0}}
\caption
 {$\delta z=(z_{phot}-z_{spec})$ against DEEP2 $B-I$ colour. The symbols are 
 the same as in Figure   \ref{fig_zz}. 
 }\label{fig_dz_colour}
\end{figure}

Figure \ref{fig_zz} compares the photometric and spectroscopic
redshift estimates for the sources with available spectroscopic
observations. With the exception of broad-line QSOs, for 
which galaxy templates are inappropriate for photometric redshift
estimates, there is fair agreement between $z_{spec}$ and $z_{phot}$
with an rms scatter, after excluding broad-line AGN, $\frac{1}{N} \sum
\left( \frac{z_{{\rm phot}} -  z_{{\rm spec}}}{1+z_{{\rm spec}}} 
\right) ^2 =0.08$, where $N$ is the total number of sources.  This is
further demonstrated in Figure \ref{fig_dz_colour} plotting $\delta
z=(z_{phot}-z_{spec})$ against $B-I$ colour. Sources bluer than
$B-I\approx1.5$ are dominated by broad-line QSOs and show significant
scatter in their photometric redshift determination (e.g. Barger et
al. 2003; Kitsionas et al. 2005). For systems redder than this limit
however, the photometric redshifts are more reliable. To avoid
erroneous redshift estimates in the analysis that follows we use
photometric redshifts only for sources with $B-I>1.5$\,mag.   

In addition to the standard photometric redshift estimation above, we
exploit the Lyman Break galaxy selection available for the GWS
(Steidel et al. 2003, 2004) to determine the redshift of a small
number of X-ray sources. This method has been shown to be very
efficient in identifying galaxies in well defined narrow redshift
slices within the range $1.5\la z \la 3$ with a low interloper
rate. Here we use the BM, BX, C, D, M and MD  Lyman Break galaxy
selection criteria fully described in Steidel et al. (2003, 2004), 
which correspond respectively to redshifts $1.70\pm0.34$,
$2.20\pm0.34$, $3.09\pm0.22$, $2.93\pm0.26$, $3.15\pm0.24$ and
$2.79\pm0.27$. In our sample there are 7 X-ray sources that
fullfill one of the above selection criteria: 2 BX, 2 MD, 1 C, 1 D and
1 M.   

\begin{figure}
\centerline{\psfig{figure=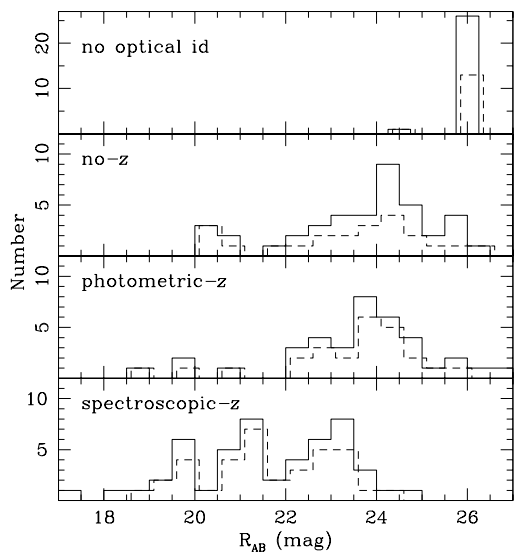,width=3in,angle=0}}
\caption
 {$R_{AB}$ optical magnitude distribution for different subsamples of 
 X-ray sources. The four panels from bottom to top plot respectively
 histograms for sources with spectroscopic redshifts, photometric
 redshifts only, without any redshift determination (but with an
 optical ID) and with no optical identification. Within each panel the
 continuous line corresponds to the full-band sample and the dashed
 line (offset to the right by 0.1\,mag for clarity) represents the
 hard sample.  
 }\label{fig_mag_z_hist}
\end{figure}

The optical spectroscopic and photometric redshift information for the
GWS sample is presented in Table \ref{tab_optical}. For the full and
hard band samples Figure \ref{fig_mag_z_hist} presents the optical
magnitude distribution of sources with spectroscopic, photometric or
no redshift information as well as sources without optical
identification. The total number of sources in these groups for
different X-ray selected samples is also shown in Table
\ref{tab_ids}. The no-redshift class involves sources that are either
too faint to estimate photometric redshifts, do not have CFHTLS data
(e.g. CCD  gaps) or have counterparts with $B-I<1.5$\,mag  and
therefore unreliable redshift determination.  

\begin{table*}
\footnotesize
\begin{center}
\begin{tabular}{l ccc ccc}
\hline
X-ray      & total  & optical & spectro-$z$ & photo-$z$ & optical ID, &  no-ID  \\
sub-sample & number & IDs     & only        &  only     & no-$z$     &\\
 \hline
 Total   &  155   &  128    &  51         & 36    &  42 &  27 \\
 Soft    &  121   &  102    &  42         & 27    &  33 &  19 \\
 Hard    &  97    &  83     &  37         & 24    &  22 &  14 \\
 Ultra-hard &  44 &  40     &  23         & 10    &   7 &   4 \\  

\hline
\end{tabular}
\end{center}
\caption{Optical and spectroscopic identification statistics. The
  columns are: (1): X-ray sub-sample; (2) total number of X-ray
  sources; (3) number of optical counterparts; (4) number of
  spectroscopic identifications; (5) number of sources with photometric
  redshift determination only (i.e. not spectroscopic $z$); (6) number
  of sources with optical counterparts for which photometric redshift
  estimations was not possible; (7) number of blank
  fields.}\label{tab_ids}  
\end{table*}

\section{X-ray spectra}\label{sec_xray_spec}

For the X-ray spectral analysis we use the {\sc xspec} v11.3.1
package. The X-ray counts of each source are extracted using the 95 per
cent encircled energy radius (1.5\,keV) at the position of the
source. The background is estimated  using an annulus centered on the
source with inner aperture size 1.5 times larger than the 95 per cent
encircled energy radius and outer aperture 100 pixel greater. We fit
the data adopting a power-law model absorbed by both an intrinsic
column density at the redshift of the source and a Galactic column at
$z=0$ fixed to $N_H=1.3 \rm \times 10^{20} \, cm^{-2}$, appropriate
for the EGS ({\sc wabs*zwabs*pow}). For the absorption we adopt the
Wisconsin cross-sections (Morrison and McCammon 1983). For sources
without spectroscopic or photometric redshift estimates we assume
$z=1.5$ to estimate the intrinsic $N_H$. Adopting a different mean $z$
in the range $1-2$ for these systems however, does not significantly
modify our results and conclusions.

In the case of sources with small number of net counts ($\la 200$) we
use the C-statistic technique (Cash 1979) specifically developed to
extract information from low signal-to-noise ratio spectra. The data are
grouped to have at least one count per bin. We attempt to
constrain the intrinsic $\rm N_H$ by fixing  the power-law index to
$\Gamma=1.8$. This value of $\Gamma$ is selected to be inbetween the
mean spectral index of radio loud ($\Gamma=1.6$; Reeves \& Turner
2000; Gambill 2003) and radio quiet AGNs ($\Gamma\approx1.9$; Nandra
\& Pounds 1994; Reeves \& Turner 2000). Adopting a single fixed $\Gamma$ 
for the spectral analysis is an approximation. We caution that our
analysis will overestimate the $N_H$ for sources with spectra
intrinsically flatter than $\Gamma=1.8$. 

For sources with sufficient number of counts ($\ga 200$) we perform
standard $\chi^{2}$ spectral fitting. The data were  grouped to have a
minimum of 20 counts per bin to ensure that Gaussian statistics
apply. For the $\chi^{2}$ analysis  we require that the source
spectrum has at least 10 spectral bins. The {\sc wabs*zwabs*pow} model 
provides acceptable fits (i.e. reduced $\chi^{2}\approx1$) for all
sources. The parameters estimated from the C-statistic and the
$\chi^{2}$ analysis are consistent within the errors.   

For both the $\chi^{2}$ and the  C-statistic analysis the fit was
performed in the 0.5-8\,keV energy range where the sensitivity of the
{\it Chandra} is the highest. The estimated errors correspond to
the 90 per cent confidence level.  The results of the X-ray spectral
analysis are presented in Table \ref{tab_optical}.

\section{The model}\label{sec_model}
In this section we describe the model we use to interpret the optical
and X-ray properties of the GWS X-ray sources in the context of AGN
evolution scenarios and different parameterisations for the intrinsic
$N_H$ distribution. Modeling of the data requires certain assumptions
about the X-ray spectra of AGN, their luminosity function and its
evolution with redshift as well as  the relative fraction of obscured
and unobscured systems. 

We model the X-ray spectra of AGN adopting for simplicity an absorbed
power-law spectral energy distribution with fixed exponent
$\Gamma=1.8$ and photoelectric absorption cross sections as described
by Morrsion \& McCammon (1983) for solar metallicity. 

For the X-ray luminosity function (XLF) of AGNs and its evolution with 
redshift we use the two different parameterisations presented by Miyaji et
al. (2000) and Ueda et al. (2003). 

Miyaji et al. (2000) combined deep pencil-beam and shallow wide-area
{\it ROSAT} surveys to estimate the XLF of
unobscured AGNs in the rest-frame 0.5-2\,keV energy band. We adopt the
luminosity dependent density evolution parameterisation of the XLF
proposed by  Miyaji et al. (their model LDDE1). These authors argue
that this model provides a better description of the observations
compared to the pure density or luminosity evolution. In this picture 
AGNs evolve differentially with more luminous systems evolving faster
than less luminous ones. Such a trend has also been proposed for the
optical luminosity function of QSOs (e.g. Wisotzki 1998).   

Ueda et al. (2003) estimated the AGN XLF in the rest-frame 2-10\,keV
energy range using a combination of hard-band ($>2$\,keV) surveys
conducted with {\it HEAO-1}, {\it ASCA} and {\it Chandra}
missions. Here we adopt the luminosity dependent density evolution of
the luminosity function, which according to Ueda et al. provides a
better fit to the data. In this parameterisation the cutoff redshift,
after which the evolution of AGN stops, increases with luminosity. We 
note that this is different from the  Miyaji et al. (2000) luminosity
dependent density evolution, where it is the rate of evolution
that changes with luminosity but not the cutoff redshift.    

For the AGN $N_H$ distribution, $f(N_H)$, we experiment with different
parameterisations. The first model adopted here is the one estimated
by Ueda et al. (2003) on the basis of  observational data and for
column densities in the range $10^{20} < N_H < 10^{24} \rm \,
cm^{-2}$. The interesting feature of their functional form is that the
fraction of obscured AGNs drops with increasing luminosity. In that
sense the Ueda et al. (2003) $f(N_H)$ does not strictly follow the
unified model prescription (Antonucci 1993) where the only  parameter
determining the obscuring column density is the viewing angle to the
observer.  
We also consider a set of models where $f(N_H)$ is a step function with
a fixed ratio, ${\cal R}$, between absorbed ($ N_H>10^{22} \rm \, cm^{-2}$)
and unabsorbed ($N_H < \rm 10^{22} \, cm^{-2}$) sources. We further
assume that obscured  and unobscured AGNs are distributed uniformly in
the range  $10^{20} < N_H < 10^{22}$ and  $10^{22} < N_H < 10^{26} \rm
\, cm^{-2}$ respectively. For the obscured systems, the above
assumption is in fair agreement with the $N_H$ distribution of
Seyfert-2s estimated by Risaliti, Maiolino \& Salvati (1999). These
authors find  that about 75 per cent of their sample has $N_H>10^{23}
\rm \, cm^{-2}$ and at least 25 per cent has $N_H>10^{25} \rm \,
cm^{-2}$. This set of models are consistent with the unified scheme
with the ratio ${\cal R}$ related to the opening angle of the
torus. Similar models but with a smooth transition between obscured
and unobscured sources are presented by Treister et al. (2004).  In
this paper we use models with  ${\cal R}=1$, 2, 3 and 4. We 
note that ${\cal R}=4$ is the locally estimated value for the fraction
of obscured AGN (Maiolino \& Rieke 1995).

\section{Results}

In this section we compare the observed redshift and column density
distributions of the GWS sample against the predictions of the
models above. To minimise incompleteness uncertainties due to either
optically unidentified sources or systems without redshift
determination (spectroscopic or photometric) we perform the comparison
for the hard-band X-ray sources. The 2-10\,keV sample has indeed,
sufficient number of sources to avoid poor statistics while, about 37
per cent of them (36 out of 97) are either blank fields or do not have
a redshift estimate. Selection at the 2-10\,keV band also
provides samples that are less sensitive to obscuration and is
therefore best suited for studies on the intrinsic fraction of
obscured AGN. Finally, many groups have published results for
the 2-10\,keV spectral band and therefore, choosing this energy range
for the analysis facilitates comparison of our survey with previous 
samples. For the comparison with the models above, the AGN XLF is
integrated in the redshift interval $z=0-5$ for unobscured
luminosities in the range $L_X(\rm 2-10 \,keV)=10^{42}-10^{46} \rm \,
erg \, s^{-1}$. For each luminosity and redshift interval the
distrubition of AGNs to different columns is described by the $f(N_H)$
models discussed in the previous section. The predicted number density
of objects in each $L_X$, $N_H$ and $z$ bins are then folded through
the sensitivity map of the EGS to estimate the number of AGNs to the
flux limit of the survey.  

\subsection{The redshift distribution}

\begin{figure}
\centerline{\psfig{figure=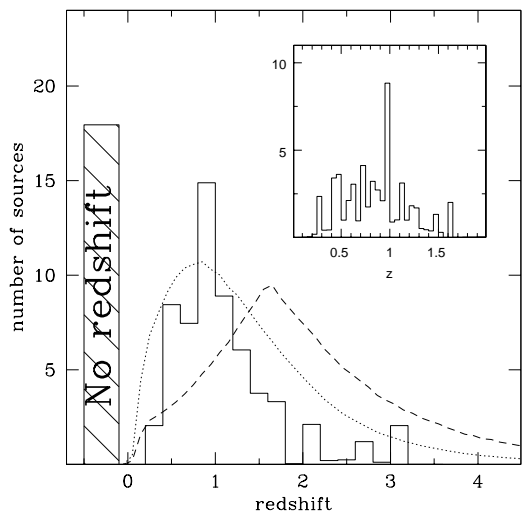,width=3in,angle=0}}
\caption
 {Redshift distribution of the hard X-ray selected sample. The hatched
 region shows the number of sources without spectroscopic
 identification. The dotted and dashed lines are the predictions of
 the Ueda et al. (2003) and Miyaji et al. (2000) XLFs
 respectively. The inlet plot shows the same redshift distribution
 only for sources in the range $0<z<2$ using a narrower bin size to
 search for cosmic structures within the surveyed area.   
 }\label{fig_z_dist}
\end{figure}

Figure \ref{fig_z_dist} shows the redshift distribution of the
hard-band sample. For sources without spectroscopic redshifts we use
the photometric redshift probability density distribution, instead of
the primary solution only, to construct the histogram in Figure
\ref{fig_z_dist}. This approach guarantees that some of the
uncertainties involved in the determination of photometric redshifts
are factored into our analysis.  For sources assigned photometric
redshifts based on the Lyman break galaxy selection (Steidel et
al. 2003, 2004) we assume a Gaussian probability density distribution
with a mean and a standard deviation appropriate for the selection
criteria that each source fulfills (see section \ref{sec_redshift}). 

A total of 36 hard X-ray selected sources (37 per cent of the sample)
do not have spectroscopic or photometric redshift determination. These
are shown with the hatched histogram  in Figure
\ref{fig_z_dist}. Their optical magnitude distribution is presented in
the two upper panels of Figure \ref{fig_mag_z_hist}. Fourteen of these
36  sources are blank fields  and are most likely associated with
$z>1$ systems. The remaining 22 sources have a distribution that is
skewed to fainter magnitudes in Figure \ref{fig_mag_z_hist} compared to
spectroscopically identified systems but similar to that of X-ray
sources with photometric redshift determination. However, 17
of these 22 sources have $B-I<1.5$\,mag and are likely to be
associated with high-$z$ QSOs. Nevertheless, unless all
spectroscopically unidentified sources are clustered in a narrow
redshift slice, we do not expect them to drastically modify the
position of the peak of the  distribution in Figure \ref{fig_z_dist}.   

Also shown in Figure \ref{fig_z_dist} are the predictions of the
two model XLFs presented in the previous section
for the Ueda et al. (2003) $f(N_H)$. The adopted $N_H$ distribution
has only minor effects on the resulting  redshift distribution and does
not affect any of our conclusions.  In Figure \ref{fig_z_dist} the
Miyaji et al. (2000) prediction peaks at $z\approx1.5$, higher than
the observations. On the contrary, the Ueda et al. (2003) XLF
produces a redshift distribution with a peak and overall 
shape in broad agreement with the data. There is however, a larger
fraction of $z\approx1$ sources compared to the model prediction,
suggesting a comsic variance spike. This is more clearly demonstrated
in the inlet plot of  Figure \ref{fig_z_dist} which uses a narrower
logarithmic redshift bin of 0.05. The full EGS sample will have
a sufficiently wide FOV ($\rm 0.5\, deg^2$) to address this issue. 
The  Ueda et al. luminosity function also predicts a larger number of
high-$z$ systems compared to the  observations. It is possible that
some the X-ray sources without redshift determination will populate
this high-$z$ tail. We attempt to quantify the agreement between the
observed and model distributions in Figure \ref{fig_z_dist}, in the
optimal case that the spectroscopically unidentified sources are
distributed to redshift bins in such a way that the difference between
the observed and model $N(z)$ is minimal. In the case of the Ueda et
al. (2003) XLF we estimate a $\chi^2$-test probability that the two
distributions (model and observations) are drawn from the same parent
population of about 99 per cent. For the  Miyaji et al. (2000) model 
this  exercise gives a probability of $<1$ per cent.

\subsection{The column density distribution}

\begin{figure}
\centerline{\psfig{figure=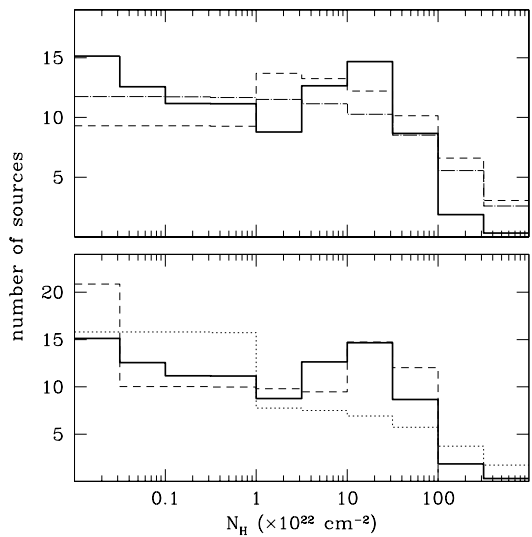,width=3in,angle=0}}
\caption
 {$N_H$ distribution of the hard X-ray selected sample. In both panels
 the bold continuous line is the observed distribution. For clarity
 the comparison with the model predictions is split into two
 panels. {\bf Upper panel:} comparison of the observations with the
 ${\cal R}$=2, 3, and 4 models described in the text corresponding to
 the dash-dotted, dashed and dotted lines respectively. {\bf Lower
 panel:} comparison of the observations with the  ${\cal R}$=1 model
 (dotted line) and the luminosity dependent $N_H$ distribution derived
 by Ueda et al. (2003; dashed line). All models are for the Ueda et
 al. (2003) XLF. 
 }\label{fig_nh_dist}
\end{figure}

Figure \ref{fig_nh_dist} presents the $N_H$ distribution of the
hard X-ray selected sample. The fraction of obscured
AGN ($\rm N_H > 10^{22} \, cm^{-2}$) in this figure is estimated
$48\pm9$ per cent. For sources without spectroscopic or
photometric redshift estimates we assume $z=1.5$ to estimate the
intrinsic $N_H$. Adopting a different mean $z$ in the range $1-2$ for
these systems however, does not significantly modify the derived $N_H$
distribution. Also, the photometric redshift uncertainties have little
impact on the observed distribution in Figure
\ref{fig_nh_dist}. Assuming a redshift dependence of the rest-frame 
column density of the form $\rm N_H \propto (1+z)^{2.65}$ (e.g. Barger
et al. 2003) and a photo-$z$ rms scatter $\delta z /  (1+z)=0.08$ (see
section 4), we estimate $\delta \log \rm N_H=0.09$, which is much
smaller than the small photon statistics uncertainty. 

Also, this plot is constructed using the $N_H$ probability density
distribution for each source instead of the best-fit solution. The
advantage of this approach is that the column density uncertainties,
due to small number of photons in the X-ray spectra, are taken into
account in the analysis. Indeed, the C-statistic, used in the few
counts limit,  provides confidence limits and probability density
distributions for the fitted parameters. In particular the $\Delta C$
is distributed as  $\chi^2$  with ${\cal \nu}$ degrees of freedom (where
${\cal \nu}$ is the number of  fitted parameters) and hence, the same
methods used to estimate confidence intervals for the model parameters
in the case of the $\chi^2$ analysis also apply to the C-statistic for
spectra with few counts (Cash 1979). We confirm this by performing
simulations of power-law spectra with fixed $\Gamma=1.8$ and different
levels of X-ray obscuration. We verify that even to the limit of less
than $\approx 10$ photons the $N_H$ probability density distribution
estimated from the C-statistic is in agreement with that derived from
the simulations.

Figure \ref{fig_nh_dist} presents the predictions of the different
$f(N_H)$ models discussed in the previous section combined with the
Ueda et al. (2003) XLF. Comparison with the data suggests that the 
${\cal R}=2$ model is in broad agreement with the observations. The
$\chi^2$-test probability that the two  distributions are drawn from
the same parent population is about 51 per cent. Higher or lower
values of  ${\cal R}$ provide poorer fits and fail to predict the
observed distribution. Using the $\chi^2$ statistical test we estimate
probabilities of 12 per cent for the ${\cal R}=3$ model and $<1$ per
cent for the ${\cal R}=1$ and 4 models. For the Ueda et
al. (2003)  luminosity dependent $f(N_H)$ we estimate a $\chi^2$-test
probability of about 71 per cent, somewhat better than the ${\cal
  R}=2$ model, the best of the step function $N_H$ distributions with
fixed ratio  ${\cal R}$. We note that the adopted XLF has little
effect on the model $\rm N_H$ distributions shown in Figure
\ref{fig_nh_dist} and therefore, does not alter our main conclusions.  

We note however, that our sample may comprise a number of Compton
thick AGN ($N_H>10^{24} \rm \, cm^{-2}$) where the direct X-ray
emission is completely blocked from view and the spectrum in the {\it
 Chandra} energy  band is a pure reflection continuum. Fitting a
single absorbed power-law to these sources is clearly not appropriate
and will produce erroneous $N_H$ estimates. We attempt to quantify
this effect by simulating reflection dominated spectra and then
fitting them with {\sc wabs*pow} {\sc xspec} models as described in section
\ref{sec_xray_spec}. For this exercise we use the Compton reflection
models of Magdziarz \& Zdziarski (1995) as implemented in the {\sc
pexrav} spectral energy distribution of {\sc xspec}. We assume a solid
angle of $2\pi$, solar abundance for all elements and an average
inclination relative to the line of sight $\cos i=0.45$. Only the 
reflection component was used, i.e. no direct radiation. We also add a
FeKa iron line assuming a Gaussian profile with width $\sigma=100\,\rm 
eV$, similar to the instrumental FWHM of the ACIS-I and rest frame
equivalent width of 1\,keV appropriate for heavily obscured AGNs
($N_H>10^{24}\rm \,cm^{-2}$; e.g. George \& Fabian 1991). For the
simulations we adopt a redshift $z=1.5$ and fix the normalisation so
that the spectrum has a flux of about $5\times10^{-15}\rm \, erg \,
s^{-1} \,cm^{-2}$ in the 2-10\,keV band. The simulated spectra are fit
with an absorbed power-law as described in section \ref{sec_xray_spec}
to estimate the $N_H$ our X-ray spectral analysis produces for this
type of sources. The simulations give a narrow distribution for the
estimated $N_H$ in the range  $10^{23}-10^{24}\, \rm cm^{-2}$. We
therefore underestimate the column density of these systems and it is
likely  that some of the sources in the range $N_H \approx
10^{23}-10^{24}\, \rm cm^{-2}$ in Figure \ref{fig_nh_dist} should be
moved to higher $N_H$ values. We note however, that even if some of
the sources in this column density range are reflection dominated
Compton thick AGN, we do not expect this to modify our conclusions
about the agreement between the data and different model $N_H$
distributions. Unfortunately, the small number of counts in most of
the obscured sources in our sample does not allow us to identify
Compton thick AGN candidates dominated by reflection emission.

\section{Discussion}

In this paper we explore the redshift and the intrinsic $N_H$
distributions of AGN using a deep 200\,ks {\it Chandra} pointing in
the Groth-Westphal Strip region. This is the first of a total of 8
observations that are currently underway as part of a deep
wide angle (0.5\,deg$^2$) X-ray survey in the Extended Groth Strip. 
A wealth of optical photometric and spectroscopic data are available
in this field (e.g. DEEP2, CFHTLS) providing optical identifications
as well as spectroscopic and photometric redshift estimates for the
X-ray population.   

The advantage of this dataset is that the detected sources are
responsible for a sizable fraction of the XRB (about 70 per cent in
the 2-10\,keV band) and have mean X-ray spectral properties consistent
with the X-ray background ($\Gamma \approx 1.4$; Nandra et al. 2005).
Nevertheless they are, on average, brighter than the extremely
faint  X-ray population identified in the ultra-deep {\it Chandra}
surveys (Alexander et al. 2003) facilitating
follow-up multi-wavelength studies. For example, the hard X-ray
selected sample comprises 97 sources of which 74 have $R<24.5$\,mag
(76 per cent) and therefore are accessible for optical spectroscopy
using 10-m class telescopes. For comparison, in the CDF-N a total of
332 sources are detected in the 2-8\,keV spectral band of which 203
(60 per cent) have $R<24.5$\,mag and 162 (50 per cent) have optical
spectroscopy available (Barger et al. 2003). 

\begin{figure}
\centerline{\psfig{figure=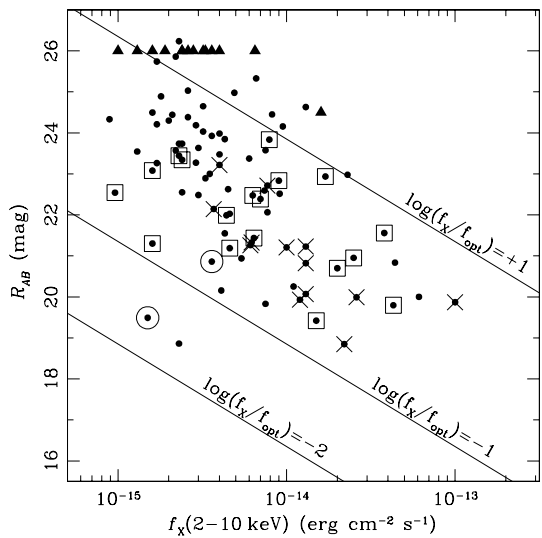,width=3.in,angle=0}}
\caption
 {$R_{AB}$-band magnitude against 2-10\,keV flux for the hard-band
 selected sample. The lines $\log f_X/f_{opt}=\pm1$ delineate the
 region of the parameter space occupied by powerful unobscured
 AGNs and are estimated from the relation $\log f_X/f_{opt} = \log
 f_X(2-10\,{\rm keV}) + 0.4\,R_{AB} +5.46$. A cross on top of a
 symbol is for sources with broad-line optical spectra. Open squares
 and open circles on top of a dot correspond to sources with narrow
 emission-line and absorption optical spectra respectively. Triangles
 represent upper limits in optical magnitude for sources without
 optical identifications.      
 }\label{fig_fxfo}
\end{figure}

\begin{figure}
\centerline{\psfig{figure=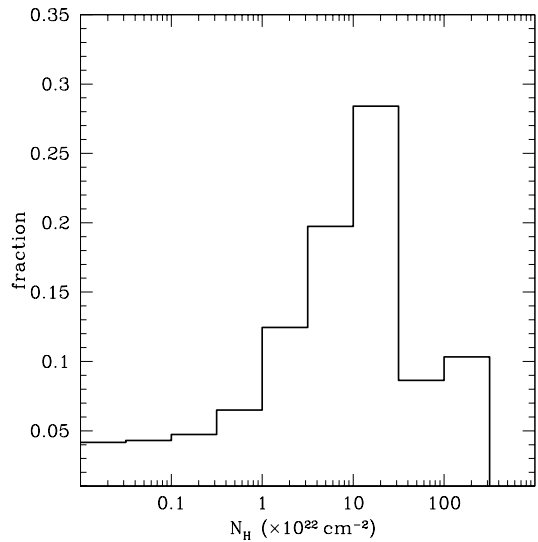,width=3in,angle=0}}
\caption
 {$N_H$ distribution of the hard X-ray selected sources with no
 optical identification. 
 }\label{fig_nh_dist_noids}
\end{figure}

For comparison with the models we focus on the hard X-ray selected
sample. This combines sufficient number of sources (97) for
statistical reliability and high optical identification rate
minimising uncertainties due to optically faint X-ray sources. A
total of 14 systems in this sample (14 per cent) are blank
fields. Many of them also have $\log f_X / f_{opt} \ga 1$. This is
shown in Figure \ref{fig_fxfo} which plots $R_{AB}$-band magnitude
against 2-10\,keV X-ray flux. High X-ray--to--optical flux ratio
systems are suggested to comprise a  large fraction of high-$z$
heavily obscured type-II QSOs (Mignoli et al. 2004;  Civano, Comastri
\& Brusa 2005). Figure \ref{fig_nh_dist_noids} plots the $N_H$
distribution of these sources   assuming a mean redshift $z=1.5$. The
optically unidentified sources in the hard sample are skewed toward
high column densities compared with identified sources, with a median
of about $\rm 5\times10^{22} \, cm^{-2}$.    

\begin{figure}
\centerline{\psfig{figure=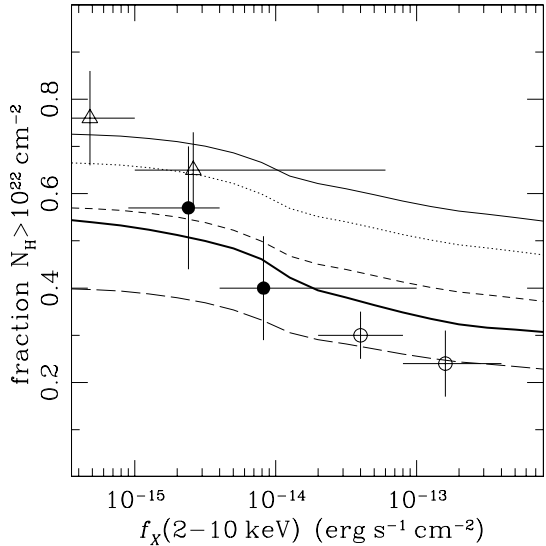,width=3in,angle=0}}
\caption
 {fraction of sources with $\rm N_H>10^{22}\,cm^{-2}$ as a function of
 2-10\,keV flux. Open triangles: Chandra Deep Field South adapted from
 Akylas et al. (2006); Open circles: the {\it XMM-Newton} survey
 described by Akylas et al. (2006); Filled circles: this study. The
 thin-line curves from bottom to top correspond to models with ${\cal
 R}=1-4$. The bold curve is the model that uses the Ueda et
 al. (2003) $N_H$ distribution.  
 }\label{fig_flux_vs_fraction}
\end{figure}

The redshift distribution of the GWS hard-band sample, using both
spectroscopic and photometric redshifts, shows a peak at $z\approx1$.
Although about 2/5 of the X-ray sources do not have redshift
determination, we argue that these systems are unlikely to modify the 
shape of the $N(z)$. The observed  redshift distribution in Figure
\ref{fig_z_dist} is in agreement with previous deep surveys in the
hard-band that also find that the X-ray population peaks at
$z\approx1$ (Fiore et al. 2003;    Georgantopoulos et al. 2004; Barger
et al. 2005; Treister et al. 2005). This is at odds with the recent
versions of the population synthesis models that successfully
reproduce the spectral properties of the XRB but predict a peak at
higher redshifts  $z\approx1.5$ (Comastri et al. 1995; Gilli et
al. 2001). As demonstrated in Figure \ref{fig_z_dist}, the origin of
this discrepancy is that the models above adopt the XLF derived for
soft-band selected powerful QSO that peak at $z\approx1.5-2$ (e.g
Miyaji et al. 2000). More recent studies however, that combine X-ray
selected AGN samples over a wider luminosity range, suggest a more
complex evolutionary history that strongly depends on $L_X$: more
powerful systems evolve up to $z\approx2$ while less luminous sources
peak at lower redshift, $z\approx1$ (Ueda et al. 2003; Barger et
al. 2005; Hasinger, Miyaji \& Schmidt 2005). As shown in Figure
\ref{fig_z_dist}, such a luminosity  dependent density evolution can
successfully reproduce the redshift distribution of our sample.  

The above complex  evolutionary pattern warrants some physical
interpretation. It is possible that less luminous systems, suggested
to  comprise a higher fraction of type-II AGN (Ueda et al. 2003;
Barger et al. 2005; Akylas, Georgantopoulos \& Georgakakis 2006), are
linked to starburst activity with a  peak at $z\approx1$, while more
powerful sources follow the QSO evolution peaking at $z\approx2$. In
this direction, the large fraction of hard (5-10\,keV) X-ray selected
sources with  mid-infrared counterparts (Fadda et al. 2002) has
motivated models where obscured X-ray sources are tied to the
infrared-luminous population with an evolution that steeply rises to
$z\approx 1$ and levels off at higher-$z$ (Franceschini, Braito \&
Fadda 2002; Gandhi \& Fabian 2003). These models produce a redshift
distribution for the faint X-ray population with a peak at $z \approx
1$ and also provide acceptable fits to both the spectral shape of the
XRB and the source counts. We note that a possible association of
X-ray obscuration and star-formation activity has also been proposed to
interpret the sub-mJy radio properties of X-ray selected AGNs (Bauer
et al. 2002; Georgakakis et al. 2004). The models above however,
require revision of the basic assumption of the unification model that
requires that type-I and II sources evolve in lockstep, since they are
drawn from the same parent population.

A key issue in the study of the origin of the XRB is the intrinsic
$N_H$ distribution of AGN. Even moderate amounts of gas ($N_H \approx
10^{22} \rm \, cm ^{-2}$) have a strong effect on the X-ray emission
below about 2\,keV and therefore this issue is better addressed by
observations at harder energies. In this respect, X-ray surveys in
the 2-10\,keV band at relatively  bright fluxes ($\ga 10^{-14} \rm \,
erg \, s^{-1} \, cm^{-2}$) suggest  that the observed fraction of AGN
with $N_H>10^{22} \rm \, cm^{-2}$ is  inconsistent with the locally
determined fraction of 4:1 (Maiolino \& Reike 1995) and closer to a
ratio of about 1:1 (Fiore et al. 2003; Perola et al. 2004;
Georgantopoulos et al. 2004; Akylas et  al. 2006). Therefore, XRB
models that adopt a fixed 4:1 fraction of obscured AGNs independent of
redshift or luminosity have a problem reproducing the observations at
bright fluxes (e.g. Treister \& Urry 2005; La Franca et al. 2005;
Akylas et al. 2006). Surprisingly, deeper
X-ray surveys show a different picture, suggesting an abrupt increase
of the fraction of obscured AGNs at faint fluxes ($\la 10^{-15} \rm \,
erg \, s^{-1} \, cm^{-2}$; Perola et al. 2004; Treister et al. 2005;
Akylas et al. 2006). This is demonstrated in Figure
\ref{fig_flux_vs_fraction} plotting X-ray flux against fraction of AGN 
with $N_H > 10^{22} \rm \, cm^{-2}$ for both the bright {\it XMM-Newton}
survey of Akylas et al. (2006) and the CDF-South. We note that for the
latter field the $N_H$ of each source is estimated using spectral
fitting instead of hardness ratio. A full description of the
spectral analysis in the CDF-South is presented by Akylas et
al. (2006). In the same figure, the results from the EGS survey, split
into two independent flux bins, are in broad agreement 
with the Akylas et al. (2006) study within the uncertainties. Also
plotted in this figure are the predictions of the different $f(N_H)$
models described in section \ref{sec_model} adopting the Ueda et
al. (2003) XLF. We note that the adopted XLF has little effect on the  
model curves plotted in Figure \ref{fig_flux_vs_fraction}. At bright
fluxes the observations are  in better agreement with ${\cal   R}
\approx 1$, while at fainter fluxes the data are progressively more
consistent with higher ${\cal R}$ models. This may suggest a
luminosity dependent $N_H$ distribution similar to that proposed by
Ueda et al. (2003). This model is indeed in fair agreement with our
data at intermediate fluxes but is not as successful at both the faint
and the bright end of Figure \ref{fig_flux_vs_fraction}, although
somewhat better than a simple fixed-${\cal R}$ $N_H$
distribution. This may suggest a steeper luminosity dependence of the
obscured AGN fraction compared to the Ueda et al. (2003)
parametrisation: i.e. $\approx  0.8$ at $L_X\approx 10^{42}\rm \, erg
\, s^{-1}$ decreasing to about $0.2$ at $L_X\approx 10^{45}\rm \, erg
\, s^{-1}$ compared to $\approx 0.6$ and 0.3 respectively for the Ueda
et al. (2003) model. We note that a number of recent studies, using
ultra-deep and/or shallow wide-angle samples, also argue in favor of 
a luminosity and/or redshift dependent $N_H$ distribution to explain
the observed properties of the X-ray population (Treister \& Urry
2005; La Franca et al. 2005; Akylas et al. 2006).

A luminosity dependent fraction of obscured AGN can be understood
in terms of modified unification schemes. In these models the  inner
radius and/or the geometric height of the torus vary with the power of
the central engine because of dust evaporation and radiation pressure
(e.g. Lawrence 1991; Simpson 2005).

\section{Concluding remarks}
This paper presents first results from an ongoing deep (200\,ks
per pointing) X-ray {\it Chandra} survey in the Extended Groth Strip 
region. Data from the first of a total of 8 pointings in this field are
used here. We analyse and discuss the optical and X-ray properties of
the sample in the context of XRB population synthesis models. 

We first construct the photometric and spectroscopic redshift
distribution of the 2-10\,keV selected sample to the limit $\approx
8 \times 10^{-15}\rm \, erg \, s^{-1} \, cm^{-2}$ which has a peak at
$z\approx1$ in agreement with previous studies. Although there is
about 40 per cent redshift incompleteness we find that luminosity
dependent density evolution, where lower luminosity systems peak at
lower redshifts, is in fair agreement with the observations. Luminosity
functions that assume evolution out to $z\approx2$ for the entire AGN
population (both type-I and II), similar to that of unobscured
broad-line QSOs do not fit the data. The  luminosity dependent
evolution supported by our data is consistent with scenarios 
suggesting that lower-luminosity systems (dominated by obscured AGNs)
are associated with star-formation activity and peak at $z \approx 1$
while, more powerful QSOs evolve out to higher-$z$. 

We also explore the $N_H$ distribution of the sample which is
consistent with either a fixed obscured AGN fraction of ${\cal
  R} \approx 2$ or the luminosity dependent $N_H$ distribution
proposed by Ueda et al. (2003), where less luminous systems comprise a 
higher fraction of type-II AGNs. We also argue that such luminosity
dependent parametrisation of the  $N_H$ distribution is essential
to account for the fraction of obscured AGN observed in different
samples over a wide range of flux limits. 

The X-ray survey of the Extended Groth Strip region, when completed,
will cover a total of about $\rm 0.5 \, deg^2$ at flux limits similar
to those presented here. The final sample will be about 8 times
larger, comprising a total of over 1000 sources. The wealth of
follow-up optical  photometric and spectroscopic data available 
in this field (DEEP2, CFHTLS) will provide optical identifications
as well as spectroscopic and photometric redshift estimates for a
large fraction of the X-ray population, for example close to 90 per
cent for the hard-band. The use of {\it Spitzer} mid-infrared data will 
further increase the number of X-ray identifications and also has the
potential to refine the photometric redshift estimates, particularly
for the optically faint subsample. Such a high identification rate
combined with photometric/spectroscopic redshifts and the
complementary multi-wavelength data (infrared, radio, sub-mm)  
available for the EGS promise a major step forward in our
understanding of the nature and the evolution of the AGN populations 
that make up the XRB.  

\section{Acknowledgments}
We thank the anonymous referee for valuable comments and suggestions. 
AG acknowledges funding from PPARC and the Marie-Curie Fellowship
grant MEIF-CT-2005-025108. This work uses data obtained with support of
the National Science Foundation grants AST 95-29028 and AST
00-71198. Funding for the DEEP2 survey has been provided by NSF grant
AST-0071048 and AST-0071198. Some of the data presented herein were
obtained at the  W.M. Keck Observatory, which is operated as a
scientific partnership  among the California Institute of Technology,
the University of California and the National Aeronautics and Space
Administration. The Observatory was made possible by the generous
financial support of the W.M. Keck Foundation. The DEEP2 team and Keck
Observatory acknowledge the very significant cultural role and
reverence that the summit of Mauna Kea has always had within the
indigenous Hawaiian community and appreciate the opportunity to
conduct observations from this mountain. 

Based on observations obtained with MegaPrime/MegaCam, a joint project
of CFHT and  CEA/DAPNIA, at the Canada-France-Hawaii Telescope (CFHT)
which is operated by the National Research Council (NRC) of Canada,
the Institut National des Sciences de l'Univers of the Centre National
de la Recherche Scientifique (CNRS) of France, and the University of
Hawaii. This work is based in part on data products produced at the
Canadian Astronomy Data Centre as part of the Canada-France-Hawaii
Telescope Legacy Survey, a collaborative project of NRC and CNRS.

\begin{table*}
\centering
\caption{Chandra GWS X-ray/optical  catalogue.
Col.(1): Source catalogue number;
Col.(2): X-ray right ascension in J2000;
Col.(3): X-ray declination in J2000;
Col.(4): X-ray/optical centroid offset in arcsec;
Col.(5): percent probability the optical counterpart is spurious alignment;
Col.(6): $R_{AB}$ magnitude and source of optical identification
         1=DEEP2, 2=Steidel et al. (2003), 3=CFHTLS;  
Col.(7): spectroscopic redshift and source catalogue 1=DEEP2, 2=DEEP
         (Weiner et al 2005), 3: CFRS (Lilly et al. 1995), 4: Steidel
         et al. (2003), 5: SDSS;
Col.(8): optical spectroscopic classification, NL=narrow emission
         lines, BL=broad emission lines, AB=absorption lines, UNCL=no
         classification available, STAR=Galactic star;
Col.(9): photometric redshift;
Col.(10): $N_H$ in units of $10^{22} \, \rm cm^{-2}$; 
Col.(11): unobscured 0.5-10\,keV X-ray luminosity in $\rm erg\,s^{-1}$.  
Col.(12); Flags fshu=source detected at
$<4 \times 10^{-6}$ probability in this band, where the bands are
f=full, s=soft, h=hard, u=ultrahard.}\label{tab_optical}
\begin{center}
\begin{tabular}{l ccc ccc ccc cc}
\hline
Cat & $\alpha_X$ & $\delta_X$ & $\delta_{OX}$ & $P$ & $R_{AB}$ &
$z_{spec}$ & class & $z_{phot}$ & $N_H$ & $L_X (\rm 0.5-10\,keV)$ & Flags \\
No. & (J2000) & (J2000) & (arcsec) & (per cent) & (mag) &   &    &   &
($\rm 10^{22}\,cm^{-2}$)  & ($10^{43}\,\rm erg\,s^{-1}$) & \\
(1) & (2) & (3) & (4) & (5) & (6) & (7) & (8) & (9) & (10) & (11) & (12) \\
\hline

c1 &14 16 42.10 & $+$52 31 42.81 & 1.36 & 0.29 & 20.95$^{1}$ &
$0.605$$^{1}$ & NL & $0.490^{+0.04}_{-0.04}$ & $0.41_{-0.28}^{+0.45}$
& 6.45 & fshu\\

c2 &14 16 43.50 & $+$52 29 02.83 & 0.00 & 0.24 & 24.42$^{3}$ & $ -
$$^{ }$ & $-$ & $-$ & $ < 2.40$ & 3.47 & f\\ 

c3 &14 16 44.03 & $+$52 30 10.40 & 0.41 & 0.12 & 22.62$^{1}$ & $ -
$$^{ }$ & $-$ & $-$ & $2.87_{-2.23}^{+3.16}$ & 8.82  & fsh\\ 

c4 &14 16 45.39 & $+$52 29 05.60 & 0.36 & 0.03 & 21.20$^{1}$ &
$1.630$$^{1}$ & BL & $0.070^{+0.1}_{-0.04}$ & $ < 1.49$ & 19.85 & fsh\\  

c5 &14 16 46.99 & $+$52 30 00.79 & $-$ & $-$ & $<$26.00$^{ }$ & $ -
$$^{ }$ & $-$ & $-$ & $4.49_{-3.17}^{+3.85}$ & 6.23 & fs\\   

c6 &14 16 48.80 & $+$52 25 58.04 & 0.73 & 0.08 & 20.84$^{1}$ & $ -
$$^{ }$ & $-$ & $-$ & $ < 0.84$ & 6.77 & fs\\ 

c7 &14 16 49.46 & $+$52 25 30.75 & 0.06 & $<0.01$ & 20.00$^{1}$ & $ -
$$^{ }$ & $-$ & $-$ & $0.17_{-0.17}^{+0.43}$ & 91.80 & fshu\\ 

c8 &14 16 51.21 & $+$52 20 47.00 & 1.35 & 0.29 & 20.70$^{1}$ &
$0.808$$^{2}$ & BL & $0.980^{+0.17}_{-0.36}$ & $ < 0.32$ & 9.48 &
fsh\\  

c9 &14 16 52.03 & $+$52 27 00.54 & 0.79 & 0.64 & 23.47$^{1}$ & $ -
$$^{ }$ & $-$ & $0.770^{+0.06}_{-0.07}$ & $2.72_{-2.06}^{+4.03}$ &
9.06 & fsh\\ 

c10 &14 16 53.46 & $+$52 21 05.54 & $-$ & $-$ & $<$24.50$^{ }$ & $ -
$$^{ }$ & $-$ & $-$ & $2.50_{-1.21}^{+1.44}$ & 23.17 & fsh\\ 

c11 &14 16 53.82 & $+$52 21 23.79 & 0.00 & 0.01 & 22.06$^{3}$ & $ -
$$^{ }$ & $-$ & $0.620^{+0.05}_{-0.06}$ & $10.05_{-4.17}^{+7.44}$ &
2.62 & fhu\\ 

c12 &14 16 58.53 & $+$52 24 12.60 & $-$ & $-$ & $<$26.00$^{ }$ & $ -
$$^{ }$ & $-$ & $-$ & $9.57_{-4.99}^{+6.96}$ & 7.59 & fsh\\ 

c13 &14 16 59.11 & $+$52 22 41.88 & $-$ & $-$ & $<$26.00$^{ }$ & $ -
$$^{ }$ & $-$ & $-$ & $28.77_{-20.29}^{+36.66}$ & 5.34  & f\\ 

c14 &14 16 59.26 & $+$52 34 36.04 & 0.00 & 1.23 & 25.03$^{3}$ & $ -
$$^{ }$ & $-$ & $-$ & $40.30_{-23.82}^{+75.73}$ & 9.18 & fh\\ 

c15 &14 17 00.03 & $+$52 23 04.41 & 0.97 & 2.09 & 24.38$^{1}$ & $ -
$$^{ }$ & $-$ & $1.270^{+0.16}_{-0.08}$ & $16.84_{-11.48}^{+89.76}$ &
3.58 & fh\\  

c16 &14 17 00.69 & $+$52 19 18.58 & 0.34 & 0.01 & 20.25$^{1}$ & $ -
$$^{ }$ & $-$ & $-$ & $ < 1.15$ & 38.42 & fsh\\
 
c17 &14 17 04.19 & $+$52 21 40.46 & 1.09 & 0.82 & 22.59$^{1}$ & $ -
$$^{ }$ & $-$ & $0.750^{+0.04}_{-0.04}$ & $7.49_{-2.25}^{+2.78}$ &
3.96 & fshu\\ 

c18 &14 17 04.26 & $+$52 24 53.78 & 0.44 & 0.01 & 19.42$^{1}$ &
$0.281$$^{1}$ & NL & $0.370^{+0.08}_{-0.08}$ & $8.82_{-2.11}^{+3.15}$
& 1.04& fshu\\

c19 &14 17 05.71 & $+$52 31 46.27 & 0.76 & 0.87 & 23.73$^{1}$ & $ -
$$^{ }$ & $-$ & $2.200^{+0.32}_{-0.32}$ & $20.58_{-19.10}^{+42.17}$ &
6.92 & fh\\  

c20 &14 17 05.75 & $+$52 32 30.62 & 0.00 & 0.07 & 22.24$^{3}$ & $ -
$$^{ }$ & $-$ & $-$ & $ < 2.02$ & 2.65 & fs\\ 

c21 &14 17 08.49 & $+$52 32 25.40 & 0.00 & 2.03 & 25.59$^{3}$ & $ -
$$^{ }$ & $-$ & $0.920^{+0.19}_{-0.18}$ & $ < 1.01$ & 0.32 & fs\\ 

c22 &14 17 08.64 & $+$52 29 29.72 & $-$ & $-$ & $<$26.00$^{ }$ & $ -
$$^{ }$ & $-$ & $1.120^{+0.13}_{-0.4}$ & $ < 1.19$ & 1.63 & fs\\ 

c23 &14 17 08.97 & $+$52 27 09.00 & 0.51 & 0.04 & 20.61$^{1}$ &
$0.532$$^{1}$ & AB & $0.540^{+0.03}_{-0.04}$ & $3.36_{-1.89}^{+2.59}$
& 0.34  & f\\ 

c24 &14 17 10.28 & $+$52 34 33.95 & 2.96 & $<0.01$ & 21.55$^{2}$ & $ -
$$^{ }$ & $-$ & $-$ & $3.99_{-1.56}^{+1.98}$ & 1.14 & fshu\\ 

c25 &14 17 10.62 & $+$52 28 28.73 & 0.24 & 0.02 & 22.02$^{1}$ & $ -
$$^{ }$ & $-$ & $-$ & $ < 0.08$ & 0.13 & fsh\\ 

c26 &14 17 11.05 & $+$52 28 37.66 & $-$ & $-$ & $<$26.00$^{ }$ & $ -
$$^{ }$ & $-$ & $-$ & $1.19_{-1.19}^{+1.86}$ & 2.89 & fshu\\ 

c27 &14 17 11.12 & $+$52 25 41.95 & 1.31 & 0.14 & 19.91$^{1}$ &
$0.418$$^{4}$ & NL & $-$ & $ < 0.08$ & 0.06 & fs\\ 

c28 &14 17 11.64 & $+$52 31 32.12 & 0.19 & 0.02 & 22.77$^{1}$ &
$0.835$$^{1}$ & NL & $0.940^{+0.33}_{-0.05}$ & $ < 0.95$ & 0.39 & fs\\

c29 &14 17 11.88 & $+$52 20 11.61 & 0.19 & $<0.01$ & 19.79$^{1}$ &
$0.433$$^{1}$ & NL & $0.600^{+0.05}_{-0.06}$ & $ < 0.04$ & 4.76 &
fshu\\  

c30 &14 17 12.90 & $+$52 22 07.53 & $-$ & $-$ & $<$26.00$^{ }$ & $ -
$$^{ }$ & $-$ & $-$ & $3.38_{-3.38}^{+20.00}$ & 1.20 & f\\ 

c31 &14 17 14.35 & $+$52 25 32.98 & $-$ & $-$ & $<$26.00$^{ }$ & $ -
$$^{ }$ & $-$ & $-$ & $1.65_{-1.65}^{+4.86}$ & 1.39 & fs\\ 
 
c32 &14 17 14.94 & $+$52 34 20.09 & 0.00 & 3.32 & 24.49$^{3}$ & $ -
$$^{ }$ & $-$ & $-$ & $26.68_{-18.28}^{+97.17}$ & 4.46 & fh\\ 

c33 &14 17 15.07 & $+$52 23 12.33 & 0.17 & $<0.01$ & 21.32$^{1}$ &
$1.263$$^{2}$ & BL & $1.700^{+0.34}_{-0.34}$ & $0.02_{-0.02}^{+0.63}$
& 8.94 & fshu\\ 

c34 &14 17 15.21 & $+$52 26 49.99 & 1.17 & 0.32 & 21.43$^{1}$ &
$0.723$$^{1}$ & NL & $0.740^{+0.07}_{-0.05}$ & $5.51_{-1.84}^{+2.15}$
& 2.23 & fshu\\ 

c35 &14 17 18.89 & $+$52 27 43.74 & 1.06 & 1.17 & 23.47$^{1}$ &
$1.211$$^{1}$ & NL & $1.120^{+0.11}_{-0.08}$ & $ < 2.15$ & 0.87 & f\\ 

c36 &14 17 19.00 & $+$52 30 51.04 & $-$ & $-$ & $<$26.00$^{ }$ & $ -
$$^{ }$ & $-$ & $-$ & $ < 3.56$ & 0.83 & fs\\ 

c37 &14 17 19.32 & $+$52 27 55.54 & 0.36 & 0.19 & 23.72$^{1}$ &
$1.208$$^{1}$ & NL & $1.130^{+0.19}_{-0.06}$ & $0.53_{-0.53}^{+3.26}$
& 0.97 & fs\\ 

c38 &14 17 20.07 & $+$52 25 00.37 & 0.10 & 0.02 & 24.03$^{1}$ & $ -
$$^{ }$ & $-$ & $0.460^{+0.09}_{-0.04}$ & $0.33_{-0.33}^{+0.54}$ &
0.20 & fsh\\ 

c39 &14 17 20.43 & $+$52 29 11.68 & 0.00 & 0.49 & 25.83$^{3}$ & $ -
$$^{ }$ & $-$ & $-$ & $ < 2.27$ & 1.31 & fs\\ 

c40 &14 17 22.98 & $+$52 31 43.50 & 0.12 & $<0.01$ & 21.30$^{1}$ &
$0.465$$^{1}$ & NL & $0.580^{+0.05}_{-0.04}$ & $ < 0.58$ & 0.24 &
fsh\\  

c41 &14 17 23.43 & $+$52 31 53.54 & 0.10 & $<0.01$ & 21.26$^{1}$ &
$0.484$$^{1}$ & BL & $0.660^{+0.06}_{-0.07}$ & $ < 0.24$ & 1.13 &
fshu\\  

c42 &14 17 23.63 & $+$52 25 55.05 & 0.38 & 0.22 & 23.54$^{1}$ & $ -
$$^{ }$ & $-$ & $-$ & $24.61_{-15.05}^{+26.83}$ & 2.08 & fh\\ 

c43 &14 17 24.30 & $+$52 32 29.61 & 0.43 & 0.19 & 23.26$^{1}$ &
$0.902$$^{1}$ & NL & $0.830^{+0.09}_{-0.04}$ & $ < 5.84$ & 0.39 & fs\\  

c44 &14 17 24.62 & $+$52 30 24.55 & 0.10 & $<0.01$ & 19.99$^{1}$ &
$0.482$$^{1}$ & BL & $0.990^{+0.17}_{-0.46}$ & $ < 0.12$ & 3.01 &
fshu\\  

c45 &14 17 25.28 & $+$52 35 12.08 & $-$ & $-$ & $<$26.00$^{ }$ & $ -
$$^{ }$ & $-$ & $0.590^{+0.57}_{-0.24}$ & $ < 0.90$ & 0.09 & f\\ 

c46 &14 17 25.37 & $+$52 35 44.19 & $-$ & $-$ & $<$26.00$^{ }$ & $ -
$$^{ }$ & $-$ & $-$ & $4.76_{-2.99}^{+4.15}$ & 6.53 & fsh\\ 

c47 &14 17 27.08 & $+$52 29 11.97 & 0.09 & 0.01 & 23.63$^{1}$ & $ -
$$^{ }$ & $-$ & $4.528^{+0.45}_{-0.45}$ & $ < 0.44$ & 5.61 & fshu\\ 

c48 &14 17 27.31 & $+$52 31 31.33 & 0.00 & 0.68 & 25.68$^{3}$ & $ -
$$^{ }$ & $-$ & $-$ & $6.71_{-6.09}^{+15.11}$ & 1.81 & fs\\ 

c49 &14 17 29.02 & $+$52 35 53.59 & 0.00 & 2.99 & 24.96$^{3}$ & $ -
$$^{ }$ & $-$ & $-$ & $0.05_{-0.05}^{+0.77}$ & 0.26 & fs\\ 

c50 &14 17 29.97 & $+$52 27 47.62 & 0.08 & 0.03 & 25.32$^{1}$ & $ -
$$^{ }$ & $-$ & $0.610^{+0.21}_{-0.07}$ & $0.29_{-0.29}^{+0.40}$ &
1.48 & fshu\\ 
\hline
\end{tabular}
\end{center}
\end{table*}

\begin{table*}
\contcaption{}
\begin{center}
\begin{tabular}{l ccc ccc ccc cc}
\hline
Cat & $\alpha_X$ & $\delta_X$ & $\delta_{OX}$ & $P$ & $R_{AB}$ &
$z_{spec}$ & class & $z_{phot}$ & $N_H$ & $L_X (\rm 0.5-10\,keV)$ & Flags \\
No. & (J2000) & (J2000) & (arcsec) & (per cent) & (mag) &   &    &   &
($\rm 10^{22}\,cm^{-2}$)  & ($10^{43}\,\rm erg\,s^{-1}$) & \\
(1) & (2) & (3) & (4) & (5) & (6) & (7) & (8) & (9) & (10) & (11) & (12) \\
\hline

c51 &14 17 30.62 & $+$52 22 42.95 & 0.39 & 0.51 & 24.79$^{1}$ & $ -
$$^{ }$ & $-$ & $1.160^{+0.12}_{-0.08}$ & $1.49_{-1.49}^{+2.93}$ &
1.51 & fs\\ 

c52 &14 17 30.66 & $+$52 23 02.21 & $-$ & $-$ & $<$26.00$^{ }$ & $ -
$$^{ }$ & $-$ & $-$ & $ < 0.89$ & 1.72 & fs\\ 

c53 &14 17 30.72 & $+$52 23 05.35 & 0.45 & $<0.01$ & 25.72$^{2}$ & $ -
$$^{ }$ & $-$ & $-$ & $ < 0.56$ & 0.50 & fs\\ 

c54 &14 17 30.87 & $+$52 28 18.22 & 0.00 & 0.79 & 25.74$^{3}$ & $ -
$$^{ }$ & $-$ & $0.550^{+0.39}_{-0.22}$ & $6.10_{-3.15}^{+5.35}$ &
0.42 & fh\\ 

c55 &14 17 32.65 & $+$52 32 02.97 & 0.12 & 0.01 & 22.83$^{1}$ &
$0.986$$^{1}$ & NL & $0.820^{+0.1}_{-0.03}$ & $0.72_{-0.66}^{+0.73}$ &
5.53 & fshu\\ 

c56 &14 17 33.64 & $+$52 20 38.86 & 2.44 & 0.96 & 20.94$^{1}$ & $ -
$$^{ }$ & $-$ & $0.530^{+0.05}_{-0.06}$ & $14.30_{-6.28}^{+14.65}$ &
1.55 & fh\\ 

c57 &14 17 33.83 & $+$52 33 49.03 & 0.16 & $<0.01$ & 21.18$^{1}$ &
$0.550$$^{1}$ & NL & $0.640^{+0.04}_{-0.07}$ & $2.16_{-0.92}^{+1.19}$
& 0.84 & fshu\\ 

c58 &14 17 34.02 & $+$52 24 56.12 & $-$ & $-$ & $<$26.00$^{ }$ & $ -
$$^{ }$ & $-$ & $-$ & $101.21_{-78.05}^{+255.55}$ & 4.46 & fh\\ 

c59 &14 17 34.41 & $+$52 31 06.63 & 0.14 & $<0.01$ & 19.49$^{1}$ &
$0.271$$^{1}$ & AB & $0.290^{+0.07}_{-0.03}$ & $0.02_{-0.02}^{+0.42}$
& 0.05 & fsh\\ 

c60 &14 17 34.87 & $+$52 28 10.45 & 0.08 & $<0.01$ & 20.82$^{1}$ &
$1.223$$^{2}$ & BL & $0.460^{+0.14}_{-0.07}$ & $ < 1.06$ & 14.08 &
fshu\\  

c61 &14 17 35.98 & $+$52 30 29.55 & 0.10 & $<0.01$ & 19.87$^{1}$ &
$0.985$$^{3}$ & BL & $0.130^{+0.12}_{-0.09}$ & $ < 0.04$ & 69.66 & fshu\\ 

c62 &14 17 36.32 & $+$52 30 16.70 & 0.14 & 0.07 & 24.62$^{1}$ &
$0.969$$^{4}$ & NL & $-$ & $3.26_{-3.26}^{+6.46}$ & 0.53 & f\\ 

c63 &14 17 36.39 & $+$52 35 44.08 & 0.49 & 0.36 & 23.81$^{1}$ & $ -
$$^{ }$ & $-$ & $0.380^{+0.07}_{-0.19}$ & $ < 0.38$ & 0.18 & fs\\ 

c64 &14 17 36.89 & $+$52 24 29.80 & 0.23 & 0.02 & 22.14$^{1}$ &
$2.125$$^{4}$ & BL & $2.056^{+0.31}_{-0.76}$ & $ < 1.04$ & 13.84
& fshu\\  

c65 &14 17 37.38 & $+$52 29 21.37 & $-$ & $-$ & $<$26.00$^{ }$ & $ -
$$^{ }$ & $-$ & $-$ & $68.09_{-36.37}^{+59.75}$ & 10.63 & fhu\\ 

c66 &14 17 38.76 & $+$52 34 13.47 & 0.63 & 0.89 & 24.18$^{1}$ & $ -
$$^{ }$ & $-$ & $-$ & $ < 1.54$ & 1.35 & fs\\ 

c67 &14 17 38.88 & $+$52 23 32.92 & 0.23 & 0.01 & 21.22$^{1}$ &
$2.148$$^{1}$ & BL & $1.691^{+0.12}_{-0.2}$ & $ < 0.69$ & 51.50 & fshu\\ 

c68 &14 17 39.06 & $+$52 28 43.78 & 0.43 & 0.65 & 24.92$^{1}$ & $ -
$$^{ }$ & $-$ & $-$ & $ < 0.01$ & 0.79 & fs\\ 

c69 &14 17 39.31 & $+$52 28 50.16 & 0.12 & 0.01 & 23.26$^{1}$ &
$0.997$$^{2}$ & UNCL & $-$ & $ < 0.01$ & 0.99  & fh\\

c70 &14 17 39.56 & $+$52 36 19.72 & 0.98 & 0.67 & 22.97$^{1}$ & $ -
$$^{ }$ & $-$ & $-$ & $2.27_{-1.95}^{+2.66}$ & 5.93 & fs\\ 

c71 &14 17 41.44 & $+$52 35 45.38 & 0.73 & 0.54 & 23.27$^{1}$ & $ -
$$^{ }$ & $-$ & $1.480^{+0.38}_{-0.38}$ & $19.64_{-8.75}^{+11.17}$ &
7.26 & fsh\\ 

c72 &14 17 41.90 & $+$52 28 23.26 & 0.10 & $<0.01$ & 21.55$^{1}$ &
$1.148$$^{1}$ & BL & $0.860^{+0.13}_{-0.03}$ & $1.16_{-0.49}^{+0.56}$
& 38.47 & fshu\\ 

c73 &14 17 42.86 & $+$52 22 35.21 & $-$ & $-$ & $<$26.00$^{ }$ & $ -
$$^{ }$ & $-$ & $-$ & $ < 0.01$ & 1.11 & s\\ 

c74 &14 17 43.28 & $+$52 20 23.10 & 0.81 & 0.01 & 17.20$^{1}$ &
$0.097$$^{5}$ & AB & $0.160^{+0.1}_{-0.03}$ & $0.15_{-0.15}^{+0.97}$ &
0.005 & fs\\ 

c75 &14 17 45.47 & $+$52 29 51.17 & 0.03 & $<0.01$ & 22.47$^{1}$ &
$0.873$$^{1}$ & NL & $0.830^{+0.07}_{-0.05}$ & $20.77_{-6.66}^{+9.40}$
& 5.66 & fshu\\ 

c76 &14 17 45.70 & $+$52 28 01.91 & 1.25 & 0.51 & 21.99$^{1}$ &
$0.432$$^{2}$ & NL & $0.400^{+0.06}_{-0.08}$ & $0.15_{-0.15}^{+0.40}$
& 0.41  & fshu\\ 

c77 &14 17 45.99 & $+$52 30 32.32 & 0.09 & $<0.01$ & 22.93$^{1}$ &
$0.985$$^{1}$ & NL & 0.$770^{+0.05}_{-0.03}$ & $3.89_{-3.46}^{+3.68}$
& 12.28 & fshu\\ 

c78 &14 17 46.17 & $+$52 25 26.65 & 0.32 & 0.10 & 23.47$^{1}$ & $ -
$$^{ }$ & $-$ & $-$ & $13.79_{-11.41}^{+84.41}$ & 2.08 & f\\ 

c79 &14 17 46.73 & $+$52 28 58.18 & 0.00 & 0.62 & 27.74$^{3}$ & $ -
$$^{ }$ & $-$ & $-$ & $3.36_{-3.36}^{+6.94}$ & 1.23 & fs\\ 

c80 &14 17 47.01 & $+$52 25 12.07 & 0.75 & 0.39 & 22.54$^{1}$ &
$0.749$$^{1}$ & NL & $0.700^{+0.05}_{-0.03}$ &
$0.05_{-0.05}^{+358.60}$ & 0.11 & fh\\ 

c81 &14 17 47.06 & $+$52 28 16.46 & $-$ & $-$ & $<$26.00$^{ }$ & $ -
$$^{ }$ & $-$ & $-$ & $1.03_{-1.03}^{+8.21}$ & 2.17 & fsh\\

c82 &14 17 47.43 & $+$52 35 10.39 & 0.65 & 0.44 & 23.21$^{1}$ &
$2.746$$^{4}$ & BL & $2.930^{+0.26}_{-0.26}$ & $0.15_{-0.15}^{+3.42}$
& 25.79 & fsh\\ 

c83 &14 17 49.21 & $+$52 28 03.28 & 0.37 & 0.14 & 23.08$^{1}$ &
$0.996$$^{1}$ & NL & $0.880^{+0.28}_{-0.02}$ & $3.63_{-3.63}^{+14.10}$
& 0.66 & fh\\ 

c84 &14 17 49.23 & $+$52 28 11.38 & 0.17 & 0.04 & 23.83$^{1}$ &
$0.998$$^{2}$ & NL & $0.830^{+0.07}_{-0.06}$ & $0.41_{-0.41}^{+0.88}$
& 6.62 & fsh\\ 

c85 &14 17 49.72 & $+$52 31 43.46 & $-$ & $-$ & $<$26.00$^{ }$ & $ -
$$^{ }$ & $-$ & $-$ & $3.04_{-3.04}^{+4.88}$ & 5.57 & fshu\\

c86 &14 17 50.19 & $+$52 36 01.15 & 0.00 & 1.68 & 26.22$^{3}$ & $ -
$$^{ }$ & $-$ & $1.100^{+0.25}_{-0.18}$ & $ < 1.44$ & 1.51 & fs\\ 

c87 &14 17 50.56 & $+$52 23 39.98 & $-$ & $-$ & $<$26.00$^{ }$ & $ -
$$^{ }$ & $-$ & $-$ & $14.51_{-12.84}^{+27.12}$ & 7.42 & fsh\\ 

c88 &14 17 50.87 & $+$52 36 32.47 & 0.57 & 0.73 & 24.18$^{1}$ & $ -
$$^{ }$ & $-$ & $-$ & $0.01_{-0.01}^{+0.44}$ & 0.01 & fs\\ 

c89 &14 17 51.00 & $+$52 25 34.13 & 0.02 & $<0.01$ & 20.86$^{1}$ &
$0.431$$^{1}$ & AB & $-$ & $21.92_{-10.75}^{+20.57}$ & 0.78 & fhu\\ 

c90 &14 17 51.18 & $+$52 23 10.96 & 0.17 & $<0.01$ & 19.83$^{1}$ & $ -
$$^{ }$ & $-$ & $0.460^{+0.03}_{-0.08}$ & $0.35_{-0.24}^{+0.35}$ &
0.74 & fshu\\ 

c91 &14 17 51.78 & $+$52 30 46.36 & 0.00 & 0.64 & 24.36$^{3}$ & $ -
$$^{ }$ & $-$ & $1.250^{+0.12}_{-0.23}$ & $2.25_{-2.25}^{+4.76}$ &
1.12 & fs\\ 

c92 &14 17 52.45 & $+$52 28 53.14 & 0.14 & 0.02 & 23.92$^{1}$ & $ -
$$^{ }$ & $-$ & $1.080^{+0.08}_{-0.08}$ & $20.46_{-16.40}^{+20.36}$ &
4.46 & fhu\\ 

c93 &14 17 52.96 & $+$52 28 38.53 & 0.43 & 0.04 & 21.39$^{1}$ &
$0.671$$^{1}$ & NL & $0.710^{+0.07}_{-0.05}$ & $ < 2.12$ & 0.07 & f\\ 

c94 &14 17 53.13 & $+$52 20 50.02 & 0.00 & 1.72 & 22.67$^{3}$ & $ -
$$^{ }$ & $-$ & $0.830^{+0.07}_{-0.05}$ & $0.19_{-0.19}^{+2.68}$ &
0.40 & f\\ 

c95 &14 17 53.72 & $+$52 34 46.34 & 0.18 & 0.01 & 22.38$^{1}$ &
$0.719$$^{1}$ & NL & $0.800^{+0.08}_{-0.06}$ &
$25.63_{-9.68}^{+22.09}$ & 4.85 & fhu\\ 

c96 &14 17 53.99 & $+$52 30 33.94 & 0.17 & 0.06 & 24.33$^{1}$ &
$0.998$$^{2}$ & UNCL & $0.830^{+0.12}_{-0.06}$ &
$4.76_{-3.15}^{+4.35}$ & 1.09 & fsh\\ 

c97 &14 17 54.25 & $+$52 31 23.37 & 0.16 & 0.06 & 24.20$^{1}$ & $ -
$$^{ }$ & $-$ & $0.810^{+0.06}_{-0.1}$ & $ < 0.26$ & 0.86 & fsh\\ 

c98 &14 17 54.58 & $+$52 34 37.95 & 0.53 & 0.29 & 23.44$^{1}$ &
$0.948$$^{1}$ & NL & $0.850^{+0.3}_{-0.04}$ &
$28.07_{-14.80}^{+35.33}$ & 2.82 & fh\\ 

c99 &14 17 55.27 & $+$52 35 32.96 & 0.35 & 0.08 & 22.55$^{1}$ &
$3.199$$^{4}$ & UNCL & $2.930^{+0.26}_{-0.26}$ &
$1.35_{-1.35}^{+8.08}$ & 16.94 & fsh\\ 

c100 &14 17 56.76 & $+$52 24 00.07 & 0.24 & $<0.01$ & 25.86$^{2}$ & $
- $$^{ }$ & $-$ & $-$ & $2.21_{-2.21}^{+3.26}$ & 3.72 & fsh\\ 
\hline
\end{tabular}
\end{center}
\end{table*}

\begin{table*}
\contcaption{}
\begin{center}
\begin{tabular}{l ccc ccc ccc cc}
\hline
Cat & $\alpha_X$ & $\delta_X$ & $\delta_{OX}$ & $P$ & $R_{AB}$ &
$z_{spec}$ & class & $z_{phot}$ & $N_H$ & $L_X (\rm 0.5-10\,keV)$ & Flags \\
No. & (J2000) & (J2000) & (arcsec) & (per cent) & (mag) &   &    &   &
($\rm 10^{22}\,cm^{-2}$)  & ($10^{43}\,\rm erg\,s^{-1}$) & \\
(1) & (2) & (3) & (4) & (5) & (6) & (7) & (8) & (9) & (10) & (11) & (12) \\
\hline

c101 &14 17 56.87 & $+$52 31 24.49 & 0.09 & 0.01 & 23.98$^{1}$ & $ -
$$^{ }$ & $-$ & $0.740^{+0.09}_{-0.07}$ & $0.01_{-0.01}^{+0.43}$ &
1.01 & fshu\\ 

c102 &14 17 56.92 & $+$52 31 18.47 & 0.00 & 0.44 & 25.10$^{3}$ & $ -
$$^{ }$ & $-$ & $-$ & $2.23_{-2.23}^{+43.75}$ & 0.53 & f\\ 

c103 &14 17 57.12 & $+$52 26 30.98 & 0.68 & $<0.01$ & 24.45$^{2}$ & $
- $$^{ }$ & $-$ & $-$ & $7.74_{-2.57}^{+3.00}$ & 14.69 & fshu\\ 

c104 &14 17 57.48 & $+$52 31 06.92 & 0.00 & 0.19 & 24.89$^{3}$ &
$3.026$$^{4}$ & UNCL & $3.150^{+0.24}_{-0.24}$ &
$20.02_{-15.41}^{+20.05}$ & 14.62 & fshu\\ 

c105 &14 17 57.51 & $+$52 25 46.45 & 0.15 & 0.02 & 23.34$^{1}$ &
$0.995$$^{1}$ & NL & $-$ & $1.29_{-1.29}^{+1.67}$ & 1.51 & fsh\\ 

c106 &14 17 58.17 & $+$52 31 33.63 & 0.00 & 1.64 & 23.56$^{3}$ & $ -
$$^{ }$ & $-$ & $-$ & $0.12_{-0.12}^{+4.13}$ & 0.42 & f\\ 

c107 &14 17 58.20 & $+$52 21 53.24 & 0.00 & 0.32 & 22.89$^{3}$ & $ -
$$^{ }$ & $-$ & $0.940^{+0.16}_{-0.08}$ & $0.65_{-0.65}^{+1.79}$ &
1.80 & fsh\\ 

c108 &14 17 58.97 & $+$52 31 38.89 & 0.20 & 0.02 & 22.71$^{1}$ &
$0.644$$^{4}$ & BL & $0.765^{+0.07}_{-0.08}$  & $0.29_{-0.29}^{+0.38}$
& 2.17 & fshu\\ 

c109 &14 17 59.32 & $+$52 24 20.30 & 0.22 & 0.07 & 23.78$^{1}$ & $ -
$$^{ }$ & $-$ & $0.320^{+0.09}_{-0.12}$ & $ < 0.41$ & 0.04 & fs\\ 

c110 &14 18 00.08 & $+$52 22 23.26 & 0.24 & 0.20 & 24.63$^{1}$ & $ -
$$^{ }$ & $-$ & $1.170^{+0.12}_{-0.13}$ & $4.29_{-4.29}^{+4.74}$ &
11.05 & fshu\\ 

c111 &14 18 00.41 & $+$52 28 22.22 & 0.61 & 0.55 & 23.73$^{1}$ & $ -
$$^{ }$ & $-$ & $-$ & $3.11_{-3.11}^{+8.05}$ &2.33 &  fh\\ 

c112 &14 18 00.43 & $+$52 36 10.11 & 0.00 & 1.16 & 26.23$^{3}$ & $ -
$$^{ }$ & $-$ & $-$ & $ < 1.56$ & 5.55 & fsh\\

c113 &14 18 01.15 & $+$52 29 41.78 & 0.06 & $<0.01$ & 23.04$^{1}$ &
$2.907$$^{4}$ & BL & $2.930^{+0.26}_{-0.26}$ & $ < 6.27$ & 5.42 & fs\\ 

c114 &14 18 01.37 & $+$52 31 50.77 & $-$ & $-$ & $<$26.00$^{ }$ & $ -
$$^{ }$ & $-$ & $1.090^{+0.1}_{-0.09}$ & $10.21_{-7.13}^{+14.04}$ &
1.20 & fh\\  

c115 &14 18 01.67 & $+$52 28 00.66 & 0.02 & $<0.01$ & 24.67$^{1}$ & $
- $$^{ }$ & $-$ & $-$ & $0.88_{-0.88}^{+5.02}$ & 0.74 & fs\\ 

c116 &14 18 02.00 & $+$52 35 14.53 & 0.32 & $<0.01$ & 19.93$^{1}$ &
$1.497$$^{1}$ & BL & $2.879^{+0.29}_{-0.31}$ & $ < 2.57$ & 111.24 &
fshu\\

c117 &14 18 02.41 & $+$52 21 32.36 & 0.00 & 0.01 & 18.86$^{3}$ & $ -
$$^{ }$ & $-$ & $0.230^{+0.09}_{-0.05}$ & $ < 0.27$ & 0.05 & fsh\\ 

c118 &14 18 02.93 & $+$52 35 47.11 & 0.00 & 0.26 & 24.97$^{3}$ & $ -
$$^{ }$ & $-$ & $-$ & $0.59_{-0.59}^{+1.42}$ & 8.09 & fshu\\ 
 
c119 &14 18 04.55 & $+$52 36 33.15 & 0.43 & 0.28 & 23.57$^{1}$ & $ -
$$^{ }$ & $-$ & $0.810^{+0.29}_{-0.04}$ & $1.01_{-0.54}^{+0.61}$ &
4.81 & fshu\\ 

c120 &14 18 04.90 & $+$52 27 40.14 & 0.18 & 0.07 & 24.30$^{1}$ & $ -
$$^{ }$ & $-$ & $0.780^{+0.05}_{-0.08}$ & $0.15_{-0.15}^{+0.68}$ &
0.79 & fsh\\ 

c121 &14 18 05.30 & $+$52 25 10.63 & 0.80 & 1.42 & 24.44$^{1}$ & $ -
$$^{ }$ & $-$ & $0.770^{+0.31}_{-0.07}$ & $0.77_{-0.77}^{+1.23}$ &
0.82 & fsh\\ 

c122 &14 18 06.51 & $+$52 33 58.67 & 0.84 & 0.72 & 23.19$^{1}$ & $ -
$$^{ }$ & $-$ & $-$ & $1.29_{-1.29}^{+4.35}$ & 1.26 & fs\\ 

c123 &14 18 07.07 & $+$52 25 23.41 & $-$ & $-$ & $<$26.00$^{ }$ & $ -
$$^{ }$ & $-$ & $0.750^{+0.4}_{-0.07}$ & $1.78_{-1.00}^{+1.07}$ &
2.49 & fsh\\ 

c124 &14 18 07.33 & $+$52 30 30.52 & 0.85 & 0.51 & 22.58$^{1}$ &
$0.990$$^{3}$ & NL & $0.850^{+0.08}_{-0.07}$ & $ < 13.12$ & 0.10 & s\\ 

c125 &14 18 08.06 & $+$52 27 50.36 & $-$ & $-$ & $<$26.00$^{ }$ & $ -
$$^{ }$ & $-$ & $-$ & $ < 12.52$ & 1.09 & f\\ 

c126 &14 18 08.94 & $+$52 31 50.84 & $-$ & $-$ & $<$26.00$^{ }$ & $ -
$$^{ }$ & $-$ & $-$ & $339.29_{-196.63}^{+271.69}$ & 34.55 & f\\ 

c127 &14 18 09.12 & $+$52 28 04.04 & $-$ & $-$ & $<$26.00$^{ }$ & $ -
$$^{ }$ & $-$ & $-$ & $6.71_{-3.67}^{+6.10}$ & 6.98 & fshu\\ 

c128 &14 18 11.26 & $+$52 30 11.48 & 0.86 & 1.12 & 23.79$^{1}$ &
$2.910$$^{4}$ & UNCL & $3.090^{+0.22}_{-0.22}$ &
$15.47_{-15.47}^{+45.56}$ &6.54 &  fs\\ 

c129 &14 18 12.16 & $+$52 28 00.29 & $-$ & $-$ & $<$26.00$^{ }$ & $ -
$$^{ }$ & $-$ & $-$ & $4.00_{-3.29}^{+4.26}$ & 3.47 & fs\\ 

c130 &14 18 13.19 & $+$52 31 13.47 & $-$ & $-$ & $<$26.00$^{ }$ & $ -
$$^{ }$ & $-$ & $0.950^{+0.27}_{-0.07}$ & $0.68_{-0.68}^{+1.15}$ &
1.76& fsh\\ 
 
c131 &14 18 13.33 & $+$52 24 14.90 & 0.00 & 1.45 & 26.81$^{3}$ & $ -
$$^{ }$ & $-$ & $1.230^{+0.15}_{-0.08}$ & $1.07_{-1.07}^{+3.37}$ &
1.53 & fs\\ 

c132 &14 18 13.96 & $+$52 26 24.79 & $-$ & $-$ & $<$26.00$^{ }$ & $ -
$$^{ }$ & $-$ & $-$ & $23.97_{-15.99}^{+24.61}$ & 5.17 & fh\\ 

c133 &14 18 14.27 & $+$52 28 10.99 & 1.02 & 0.49 & 22.22$^{1}$ &
$2.818$$^{4}$ & BL & $2.552^{+0.25}_{-0.26}$ & $ < 1.65$ & 8.47 & fs\\ 

c134 &14 18 15.36 & $+$52 32 47.61 & 0.00 & 1.29 & 24.28$^{3}$ & $ -
$$^{ }$ & $-$ & $-$ & $0.87_{-0.87}^{+1.57}$ & 0.93 & fs\\ 

c135 &14 18 16.29 & $+$52 29 40.30 & 0.18 & $<0.01$ & 20.07$^{1}$ &
$1.603$$^{3}$ & BL & $1.700^{+0.34}_{-0.34}$ & $ < 0.27$ & 34.39 &
fshu\\

c136 &14 18 16.35 & $+$52 25 24.05 & 0.61 & 0.83 & 24.18$^{1}$ & $ -
$$^{ }$ & $-$ & $-$ & $0.58_{-0.58}^{+1.45}$ & 3.47 & fsh\\ 

c137 &14 18 16.43 & $+$52 33 29.77 & 0.12 & 0.03 & 24.22$^{1}$ & $ -
$$^{ }$ & $-$ & $-$ & $ < 1.53$ & 0.29 & fs\\ 

c138 &14 18 16.73 & $+$52 23 07.98 & 0.75 & 0.57 & 23.37$^{1}$ & $ -
$$^{ }$ & $-$ & $-$ & $ < 0.32$ & 1.23 & fsh\\ 

c139 &14 18 18.04 & $+$52 32 01.28 & 0.99 & 3.31 & 24.64$^{1}$ & $ -
$$^{ }$ & $-$ & $0.390^{+0.07}_{-0.12}$ & $1.44_{-0.99}^{+1.25}$ &
0.17 & fs\\ 

c140 &14 18 19.92 & $+$52 21 15.80 & $-$ & $-$ & $<$26.00$^{ }$ & $ -
$$^{ }$ & $-$ & $1.062^{+0.11}_{-0.12}$ & $9.76_{-7.03}^{+16.59}$ &
2.85 & f\\ 

c141 &14 18 20.30 & $+$52 33 51.08 & 1.85 & 3.51 & 23.15$^{1}$ & $ -
$$^{ }$ & $-$ & $2.200^{+0.32}_{-0.32}$ & $0.02_{-0.02}^{+4.52}$ &
5.97 & fs\\ 

c142 &14 18 21.37 & $+$52 26 55.46 & $-$ & $-$ & $<$26.00$^{ }$ & $ -
$$^{ }$ & $-$ & $-$ & $ < 5.18$ & 1.83 & fs\\ 

c143 &14 18 21.39 & $+$52 32 54.31 & 0.44 & 0.29 & 23.57$^{1}$ & $ -
$$^{ }$ & $-$ & $-$ & $3.58_{-3.22}^{+4.28}$ & 0.82 & fsh\\ 

c144 &14 18 21.79 & $+$52 29 55.82 & 0.90 & 0.06 & 19.71$^{1}$ &
$0.000$$^{3}$ & STAR & $-$ & $-$ & $-$ & fs\\ 

c145 &14 18 22.08 & $+$52 26 50.20 & 0.55 & 0.14 & 22.06$^{1}$ & $ -
$$^{ }$ & $-$ & $0.740^{+0.06}_{-0.04}$ & $0.60_{-0.60}^{+1.31}$ &
0.64 & fs\\ 

c146 &14 18 22.41 & $+$52 36 07.45 & 1.31 & 3.83 & 24.15$^{1}$ & $ -
$$^{ }$ & $-$ & $-$ & $0.57_{-0.57}^{+1.13}$ & 15.50 & fshu\\ 

c147 &14 18 22.84 & $+$52 27 10.11 & 1.54 & 0.07 & 18.26$^{1}$ &
$0.281$$^{5}$ & AB & $0.530^{+0.04}_{-0.05}$ & $ < 0.08$ & 0.04 & fs\\ 

c148 &14 18 23.07 & $+$52 21 14.50 & 0.00 & 0.73 & 23.84$^{3}$ & $ -
$$^{ }$ & $-$ & $1.050^{+0.06}_{-0.07}$ & $10.50_{-6.21}^{+7.59}$ &
4.81 & fh\\ 

c149 &14 18 24.97 & $+$52 23 30.55 & 0.66 & 0.31 & 22.97$^{1}$ & $ -
$$^{ }$ & $-$ & $1.040^{+0.16}_{-0.08}$ & $3.48_{-2.10}^{+2.17}$ &
17.92 & fshu\\ 

c150 &14 18 25.52 & $+$52 23 49.48 & 0.10 & $<0.01$ & 22.51$^{1}$ & $
- $$^{ }$ & $-$ & $-$ & $0.31_{-0.31}^{+0.91}$ & 2.52 & fshu\\ 

c151 &14 18 26.36 & $+$52 28 18.80 & 0.81 & 0.68 & 23.00$^{1}$ & $ -
$$^{ }$ & $-$ & $-$ & $ < 0.79$ & 7.16 & fsh\\ 

c152 &14 18 26.44 & $+$52 32 35.01 & 0.71 & 0.04 & 19.91$^{1}$ & $ -
$$^{ }$ & $-$ & $0.730^{+0.01}_{-0.01}$ & $ < 0.27$ & 0.33 & fs\\ 

c153 &14 18 26.51 & $+$52 25 59.70 & 0.98 & $<0.01$ & 24.65$^{2}$ & $
- $$^{ }$ & $-$ & $-$ & $12.00_{-6.92}^{+10.85}$ & 5.89 & fsh\\ 

c154 &14 18 29.76 & $+$52 27 09.39 & 0.56 & 0.15 & 22.48$^{1}$ & $ -
$$^{ }$ & $-$ & $0.720^{+0.05}_{-0.03}$ & $6.36_{-3.18}^{+4.97}$ &
1.30 & fh\\ 

c155 &14 18 30.24 & $+$52 22 12.14 & 0.16 & $<0.01$ & 20.83$^{1}$ & $
- $$^{ }$ & $-$ & $-$ & $ < 0.13$ & 107.14& fshu\\ 

c156 &14 18 32.87 & $+$52 23 49.48 & 0.74 & 0.06 & 20.15$^{1}$ & $ -
$$^{ }$ & $-$ & $-$ & $ < 2.33$ & 29.17 & fsh\\ 

c157 &14 18 37.96 & $+$52 20 34.62 & $-$ & $-$ & $<$26.00$^{ }$ & $ -
$$^{ }$ & $-$ & $-$ & $ < 1.52$ & 2.53& s\\ 

c158 &14 18 38.18 & $+$52 23 58.55 & 1.99 & 0.18 & 18.84$^{1}$ &
$1.118$$^{5}$ & BL & $1.500^{+0.15}_{-0.17}$ & $ < 0.21$ & 17.79 &
fsh\\

\hline
\end{tabular}
\end{center}
\end{table*}

\end{document}